\documentclass[prb,aps,epsf,twocolumn,superscriptaddress]{revtex4-1}
\usepackage{graphicx,amsfonts,times,bm,amsmath,verbatim,color,array}

\usepackage{subfigure}
\usepackage{epsfig}
\usepackage{hyperref}
\usepackage{bm,wrapfig}
\hypersetup{
colorlinks=true,final=true,
        linkcolor=blue,
        citecolor=blue,
        filecolor=blue,
        urlcolor=blue,
}
\usepackage{graphicx}
\usepackage{epsfig}
\usepackage{amsfonts}
\usepackage{amsmath}
\usepackage{amssymb}
\usepackage{color}

\newcommand\ba{\begin{eqnarray}}
\newcommand\ea{\end{eqnarray}}
\newcommand\be{\begin{equation}}
\newcommand\ee{\end{equation}}
\newcommand\bi{\bibitem}

\def\non{\nonumber}

\def\al{\alpha}

\def\si{\sigma}

\begin{document}
\title{Magneto-transport phenomena of type-I multi-Weyl semimetals in co-planar setups}

\author{Tanay Nag}
\affiliation{Max-Planck Institute for the Physics of Complex Systems, D-01187 Dresden, Germany}
\author{Snehasish Nandy}
\affiliation{Max-Planck Institute for the Physics of Complex Systems, D-01187 Dresden, Germany}
\affiliation{Department of Physics, Indian Institute of Technology Kharagpur, W.B. 721302, India}

\begin{abstract}

Having  the chiral anomaly induced magneto-transport phenomena extensively studied in single Weyl semimetal (WSM) 
as characterized by topological charge  $n=1$,
we here address the transport properties in the context of multi-Weyl semimetals (m-WSMs) where 
$n>1$. Using semiclassical Boltzmann transport formalism
with the relaxation time approximation, we investigate
several intriguing transport properties such as longitudinal magneto-conductivity (LMC), 
planar Hall {conductivity} (PHC),
thermo-electric coefficients (TECs) and planar Nernst coefficient (PNC) \textcolor{black}{ for m-WSMs in the  
co-planar setups with {external} magnetic field, electric
field and temperature gradient}.
Starting from the \textcolor{black}{low-energy model}, we show analytically that 
at zero temperature both
LMC and PHC vary cubically with topological charge as $n^3$ while the finite temperature ($T
\neq 0$) correction is 
proportional to $(n+n^2)T^2$. Interestingly,  
{we find 
that both the longitudinal and transverse TECs} vary quadratically with topological charge as $n^2$ 
and the PNC is found to vary non-monotonically as a function of $n$.
\textcolor{black}{Our study hence clearly suggests that the inherent properties of m-WSMs indeed show up distinctly through the chiral anomaly and the chiral magnetic effect induced transport coefficients in two different setups.} 
Moreover, in order to obtain an experimentally realizable picture, 
 we simultaneously verify our analytical findings  through the numerical calculations {using} the 
 lattice model of m-WSMs.

\end{abstract}
\maketitle

\section{Introduction}

In the field of three-dimensional (3D) topological systems, Weyl semimetal (WSM) has emerged as a prime topic of interest.
In condensed matter physics, Weyl fermion appears as a low energy excitation of gapless chiral fermion near the touching of a
pair of non-degenerate
bands~\cite{Murakami_2007, Peskin_1995, Murakami2:2007, Yang:2011, Burkov1:2011, Burkov:2011, Volovik, Wan_2011, Xu:2011}.
The non-trivial topological properties of the WSMs appear due to Weyl nodes. The Weyl node describing the singularity in k-space 
acts as a source or sink of the Berry curvature. According to no-go theorem, the Weyl nodes always come in pairs of positive and 
negative topological charges (also referred to as chirality) and total topological charge in the Brillouin zone vanishes~\cite{Nielsen:1981, Nielsen:1983}. 
In order to have a topological charge (designated by $n$)  associated with the Weyl node, WSM has to break either time-reversal symmetry (TRS) or the space inversion symmetry (IS)~\cite{Burkov:2011, Volovik, Wan_2011, Xu:2011}. 
The topological charge whose strength is related to the Chern number 
 is quantized to integer values~\cite{Xiao_2010}.

The WSM phase has been realized experimentally in several inversion asymmetric compounds 
(TaAs, MoTe$_{2}$, WTe$_{2}$) without breaking TRS~\cite{Lv_2015, Huang_2015, Hasan_2015, Wu_2016, Jiang_2017,Yan_2017}. 
However, all of these materials mentioned above belong to single Weyl semimetal, whose energy dispersions are 
linear in wave vectors and topological charge equals to $\pm 1$. Recently, it has been proposed that the multi-Weyl fermions 
can also be realized in condensed matter systems~\cite{Xu:2011,bernevig12, hasan16, Nagaosa_2014}. The multi-Weyl semimetals (m-WSMs) are referred to those
materials which contain Weyl nodes with topological charge higher than 1 (i.e. $n>1$). The quasi-particle dispersion for $n> 1$ 
shows natural anisotropy in dispersion. The double WSM ($n = 2$) and triple WSM ($n = 3$) show linear dispersion along one
symmetry direction and quadratic and cubic energy dispersion relations for the other two directions respectively. From the density functional theory (DFT)
calculations, it has been suggested that HgCr$_2$Se$_4$ and SrSi$_2$ can be the candidate 
materials for double WSM~\cite{Xu:2011,bernevig12, hasan16} whereas \textcolor{black}{A(MoX)$_3$ (with $A=Rb$, $TI$; $X=Te$) kind of materials can accommodate triple-Weyl points~\cite{zunger17}}. Discrete rotational symmetry
in a lattice imposes a strict restriction that only the Weyl nodes with topological charge $n \leq 3$ can be permitted 
in real materials~\cite{bernevig12,Nagaosa_2014}.  Moreover, the single WSM can be viewed as 3D analogue of graphene
whereas the double WSM and triple WSM can be represented as 3D counterparts of bilayer~\cite{falko06} and ABC-stacked
trilayer graphene~\cite{peres06,macdonald08}, respectively.

The single WSM exhibits several fascinating transport properties in the presence as well as the absence of 
external fields. Negative longitudinal magnetoresistance (LMR) and planar Hall effect (PHE) are the two most important
transport properties which appear due to the non-conservation of separate electron numbers of opposite chirality for 
relativistic massless fermions, an effect known as the chiral or Adler-Bell-Jackiw
anomaly~\cite{Goswami:2013, Adler:1969, Bell:1969, Nielsen:1981, Nielsen:1983, Aji:2012, Zyuzin:2012, Volovik, Wan_2011, Xu:2011,Moore_2015}. \textcolor{black}{This is in contrast to the chiral magnetic effect (CME) which refers to an electric current flowing along the direction of the applied magnetic field triggered by the chirality imbalance in the Weyl nodes without any electric field}.
In recent years, these magneto-transport properties in Dirac and Weyl SM are extensively studied both
theoretically and experimentally~\cite{Kim:2014, Son:2013, Fiete_2014, He:2014, Liang:2015,CLZhang:2016,QLi:2016, Xiong, Hirsch,Sharma:2016, Tewari_2017, Vladimir_2017, Ma_2019, Burkov_jpcm, Pavan_2013, Burkov_2017, Nandy_2017, Nandy_2018, Spivak_2016, Das_2018, Yip_2015, Jia_2016, Xu_2016, Erfu_2016, Li_2018, Liang_2018, Wang_2018, deng19, Chen_2018, Singha_2018,
kumar18}. \textcolor{black}{It is noteworthy that 
{the chiral anomaly (CA) induced} PHE, observed for a 
coplanar arrangement of electric and magnetic fields, is characteristically different from
{the Lorentz force mediated} conventional Hall effect where transverse arrangement between the
above fields is {required}.}
\textcolor{black}{Although, the transport properties in the presence as well as  absence of external magnetic field have recently been studied in m-WSMs using both the low-energy model and the lattice model~\cite{Roy_2016,park_2017,Roy_2018,Gorbar_2017,Gorbar_2018,Rodrigo_2020,Roy_2020,
Wang_2019,Sengupta_2019,Liu_2018,Fiete_2016}, planar Hall conductivity (PHC) has not been studied in m-WSM so far.} In particular, the effects of enhancement of the density of states,
anisotropic nonlinear energy dispersion, and modified spin-momentum locking structure on PHE
in m-WSMs {remain unexplored.}

\textcolor{black}{The thermo-electric phenomena such as Peltier coefficient, Nernst effect and longitudinal magneto-thermal conductivity are well studied in the context of regular Dirac and Weyl SMs using semiclassical Boltzmann theory~\cite{Vladimir_2017,Sharma:2016,sharma17,Fiete_2014, 
Zyuzin_2017,Saha_2018,
Chernodub_2018, Spivak_2016,Trivedi_2017,Nandy1_2017} and are also recently observed in experiments~\cite{Watzman_2018,Liang_2017,Rana_2018,Hess_2018}. Moreover, the thermo-electric transport properties in m-WSMs have been studied~\cite{Gorbar_2017,Gorbar_2018,Fiete_2016}. There exist a plethora of theoretical works studying mainly the anomalous or conventional Nernst response in WSMs \cite{Vladimir_2017,Sharma:2016,Fiete_2014, Spivak_2016,sharma17,
Zyuzin_2017,Saha_2018,
Chernodub_2018, Trivedi_2017,Nandy1_2017,Gorbar_2017,Gorbar_2018,Fiete_2016}. An anomalous Nernst effect requires the presence of Berry curvature in a direction perpendicular to both the applied temperature gradient and the induced voltage whereas in the case of conventional Nernst effect, external magnetic field has to be applied perpendicular to both $\nabla {\mathbf T}$ and induced voltage. In this work, we study an unconventional Nernst effect, namely, the planar Nernst effect (PNE) which is different from the conventional Nernst effect as well as the anomalous Nernst effect, is known to occur in ferromagnetic systems~\cite{Avery_2012,Pu_2006,Back_2013}. We consider a situation where both the applied thermal gradient $\mathbf{\nabla T}$ and magnetic field $\mathbf{B}$ are in-plane but not parallel to each other. This situation generates an in-plane transverse voltage and the corresponding phenomenon is referred to as the PNE. Actually, PNE is the thermal counterpart of the PHE where {the applied} electric field $\mathbf{E}$ is replaced by ${\nabla \mathbf T}$. Therefore, it is now natural question to ask how PNE behaves in m-WSMs. Moreover, in the planar Nernst setup (i.e. both the applied thermal gradient $\mathbf{\nabla T}$ and magnetic field $\mathbf{B}$ are in-plane but not parallel to each other), the response of thermo-electric
coefficients in the context of m-WSMs has not been explored yet.}

\textcolor{black}{In this paper, we study several intriguing transport coefficients in m-WSMs considering the co-planar
setups. Using the low-energy model of m-WSMs, we first analytically calculate 
longitudinal magneto-conductivity (LMC),
planar Hall conductivity (PHC), longitudinal thermo-electric
coefficient (LTEC) and transverse thermo-electric coefficient (TTEC) (usually referred to as the Peltier coefficient). Interestingly, we 
find that both LMC and PHC go as $n^3$ at zero temperature while the finite temperature correction is ${\mathcal O}((n+n^2)T^2)$. On the other hand, both LTEC and TTEC follow $n^2 T$ dependence. Moreover, we find that LMC and LTEC show $B^2 \cos^2 \gamma$ dependence whereas PHC and TTEC are proportional to  $B^2 \sin \gamma \cos \gamma$ in m-WSMs. Here, $\gamma$ is the angle between applied $\mathbf{B}$ and $\mathbf{E}$ for the 
measurement of PHE or between applied $\mathbf{B}$ and $\mathbf{\nabla T}$ for the measurement of thermo-electric coefficients. Secondly, using the thermo-electric tensor and the charge conductivity tensor, we are able to calculate the functional form of planar Nernst coefficient (PNC). We find that PNC, which is proportional to $B^2 \sin \gamma \cos \gamma$, does not show any monotonic dependence on topological charge as compared to LMC, PHC or TECs.
Finally, in order to get a complete picture and verify our analytical findings,
we numerically investigate the magnetic field dependence, and angular dependence of electrical conductivity, thermo-electric coefficients and PNE considering the lattice models of m-WSMs.}

The rest of the paper is organized as follows. In Sec.~\ref{model},
we introduce the low-energy Hamiltonian as well as TRS breaking lattice Hamiltonian for 
m-WSMs. Sec.~\ref{Boltzmann}
is devoted to the general expressions of LMC, PHC, TECs and PNC. In Sec.~\ref{result},
analytical expressions (Sec.~\ref{Lin_Results1}) using
low-energy model and numerical results (Sec.~\ref{result_Latt})
considering the lattice model of m-WSMs are presented
for \textcolor{black}{different transport properties such as LMC, PHC, TECs
and PNC. The analytical calculations are given in detail in the Appendix \ref{app1} and \ref{app2}.}
Finally, we summarize our results and discuss possible
future directions in Sec.~\ref{cons}. 

\section{Model Hamiltonian}
\label{model}
\subsection{Low-energy Hamiltonian}
The low-energy effective Hamiltonian describing the  Weyl node with topological charge $n$ can 
be written as~\cite{Xu:2011, bernevig12,Nagaosa_2014, Roy_2017} 
\be
H_{n} \left( \mathbf{k} \right) = \alpha_{n} k^n_{\bot} \left[ \cos \left( n \phi_{k} \right) 
\sigma_{x} +\sin \left( n \phi_{k} \right) \sigma_{y} \right] + v k_z \sigma_{z} 
\label{eq_multi1}
\ee
where $k_{\bot}=\sqrt{k_x^2+k_y^2}$ and $\phi_k={\rm arctan}(k_y/k_x)$. Here, $\alpha_n$ bears the connection to the Fermi velocity. For example, $\alpha_1$ has the dimension of Fermi velocity, while $\alpha_2$
has the dimension of mass. $v$ is equivalent to the velocity associated with $z$-direction.
Here, $\sigma_i$'s $\left( \sigma_x,\sigma_y,\sigma_z \right)$ are the Pauli matrices representing the pseudo-spin indices.
The Hamiltonian {given in Eq.~(\ref{eq_multi1})} can be written in a compact form as $H=\mathbf{n}_{\mathbf{k}}\cdot \boldsymbol{\sigma}$
with $\mathbf{n}_{\mathbf{k}}=( \alpha_{n} k^n_{\bot}\cos \left( n \phi_{k} \right), \alpha_{n} k^n_{\bot}\sin 
\left( n \phi_{k} \right), v k_z )$. The energy dispersion of the Weyl node is given by 
\begin{equation}
\epsilon_{\mathbf{k}}^{\pm} = \pm \sqrt{ \alpha^2_{n} k^{2 n}_{\bot} + v^2 k^2_z}  
\label{eq_multi2}
\end{equation}
where $\pm$ represents conduction and valence bands respectively. 
It is clear from the Eq.~(\ref{eq_multi2})
that the topological charge determines not only the topological nature of
the wave function but also the anisotropic energy dispersion of the system.
\textcolor{black}{The single Weyl dispersion $\epsilon_{\mathbf{k}}=v\sqrt{k^2_x+ k^2_y+ k^2_x}$ can be obtained
by setting $n=1$ and $\alpha_1=v$ in Eq.~(\ref{eq_multi2}).} Therefore, it is clear that the dispersion around a Weyl node with $n=1$
is isotropic in all momentum directions. \textcolor{black}{On the other hand, for $n>1$, we find that the dispersion 
around a double Weyl node ($n=2$) becomes quadratic along both $k_x$ and $k_y$ directions whereas varies linearly with $k_z$.
Substituting $n=3$ in Eq.~(\ref{eq_multi2}), it is easy to see that the dispersion around a triple Weyl ($n=3$) node is cubic along both $k_x$ and $k_y$
directions and becomes linear in $k_z$ direction. 
We additionally note that 
{in this study} we restrict ourselves to type-I m-WSM (Eq.~\ref{eq_multi1}) where a single multi-Weyl node, 
{separated from the opposite chirality multi-Weyl node in momentum space,}  
is presented with the absence of the tilt parameter.}

The Berry curvature of the m$^{\textrm{th}}$ band for a Bloch Hamiltonian $H_n({\mathbf k})$, defined as the Berry phase per
unit area in the ${\mathbf k}$ space, is given by ~\cite{Xiao_2010}
\begin{equation}
\Omega^{m}_{a} (\mathbf{k})= (-1)^m \frac{1}{4|n_{\mathbf{k}}|^3} \epsilon_{a b c} \mathbf{n}_{\mathbf{k}} 
\cdot \left( \frac{\partial \mathbf{n}_{\mathbf{k}}}{\partial k_b} \times \frac{\partial \mathbf{n}_{\mathbf{k}}}{\partial k_c} \right) .
\label{bc_lattice}
\end{equation}
The explicit form of different Berry curvature components associated with the
multi-Weyl node are given by
\begin{equation}
{\Omega}_{\mathbf{k}}^{\pm} =\pm \frac{1}{2} \frac{n v \alpha_n^2 k^{2n-2}_{\bot} }{\epsilon_{\mathbf{k}}^{3}}
\: \left( k_x, k_y, n k_z \right).
\label{eq_bcl}
\end{equation}

The Berry curvature of a single WSM can be easily obtained from Eq.~(\ref{eq_bcl}) by 
 setting $n=1$ and $\alpha_1=v$ which gives ${\Omega}_{\mathbf{k}}^{\pm}= \pm \mathbf{k}/\epsilon^3_{\mathbf{k}} $ with 
 $\mathbf{k}=(k_x,k_y,k_z)$. Therefore, the Berry curvature is isotropic in all momentum directions for single Weyl case. On the other hand, it is clear from Eq.~(\ref{eq_bcl}) that the Berry curvature becomes anisotropic for WSMs with $n>1$ i.e., for double WSM ($n=2$) and triple WSM ($n=3$) due to the presence of $k^{2n-2}_{\bot}$ factor and monopole charge $n$. \textcolor{black}{In particular, we find that $\Omega_z$ is algebraically dependent on $n$ in a quadratic manner while $\Omega_x$ and $\Omega_y$ bear linear algebraic dependence on $n$.}
Therefore, the multi-Weyl nature can indeed modify Berry curvature induced transport properties in double and triple WSMs as compared to single Weyl case.

\textcolor{black}{The components of the quasi-particle velocity ($v_{\mathbf{k}}=\frac{\partial \epsilon_{\mathbf{k}}}{\partial \mathbf{k}}$) associated with the multi-Weyl node are given by}
\be
v_{\mathbf{k}}=\frac{1}{\epsilon_{\mathbf{k}}}(k_x n \alpha_n^2 k_{\bot}^{2(n-1)},k_y n \alpha_n^2 k_{\bot}^{2(n-1)},v^2 k_z).
\label{eq_vel}
\ee
\textcolor{black}{It is clear from Eq.~(\ref{eq_vel}) that the velocity for a single WSM is
$v_{\mathbf{k}}=v(k_x,k_y,k_z)/\epsilon_{\mathbf{k}}$ which shows the isotropic nature of the velocity in all
momentum directions. One can figure out from the same equation that the velocity is no longer isotropic if
we consider WSMs with $n>1$ compared to the single WSM.} In particular, since the energy dispersion becomes anisotropic 
in double and triple WSMs as described in Eq.~(\ref{eq_multi2}), the $x$ and $y$ components of the velocity vary
with different power in $k_x$ and $k_y$ due to the factor $k_{\bot}^{2(n-1)}$ in these cases  while 
$v_z$ remains unaltered (varies linearly with $k_z$) irrespective of the value of $n$.

\begin{figure*}[htb]
\begin{center}
\epsfig{file=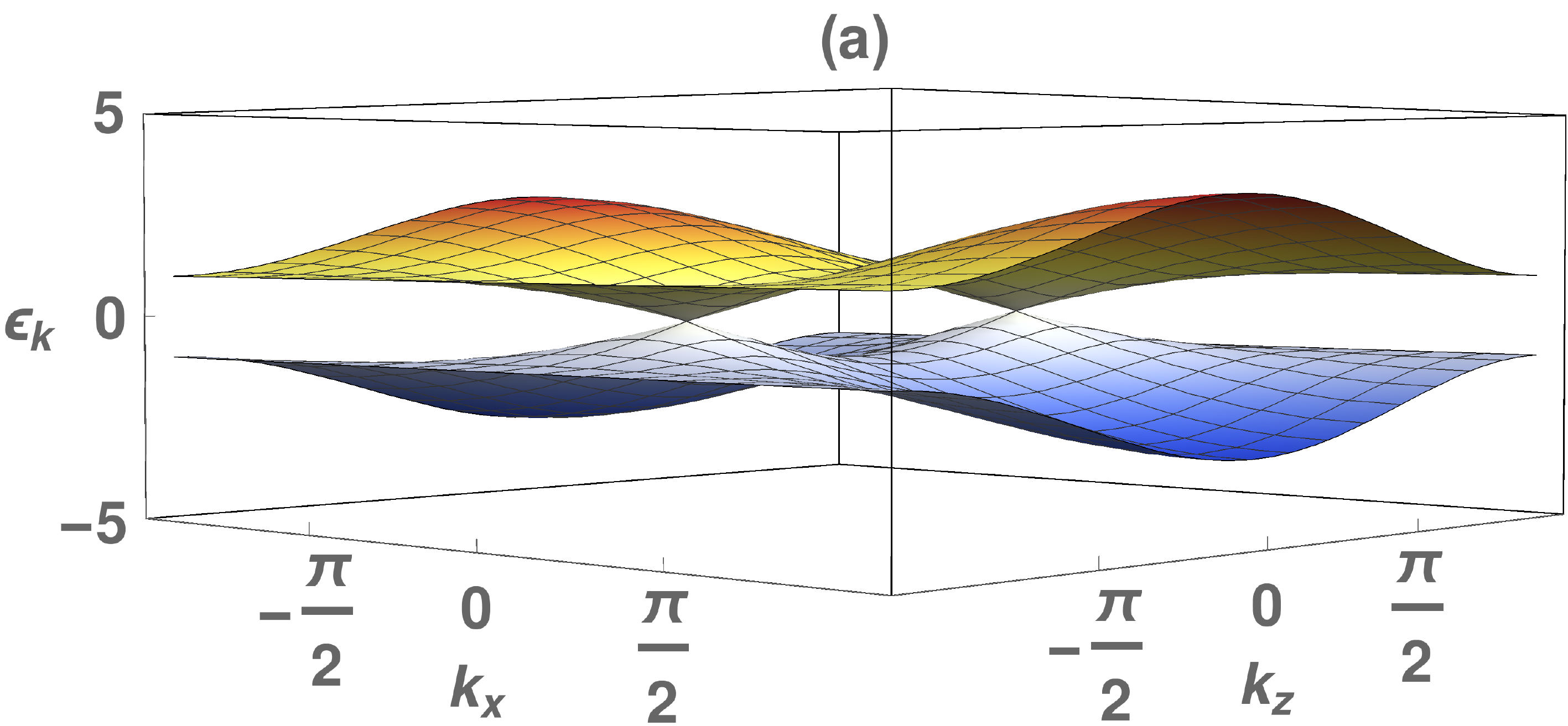,trim=0.0in 0.05in 0.0in 0.05in,clip=true, width=55mm, height=40mm}\vspace{0em}
\epsfig{file=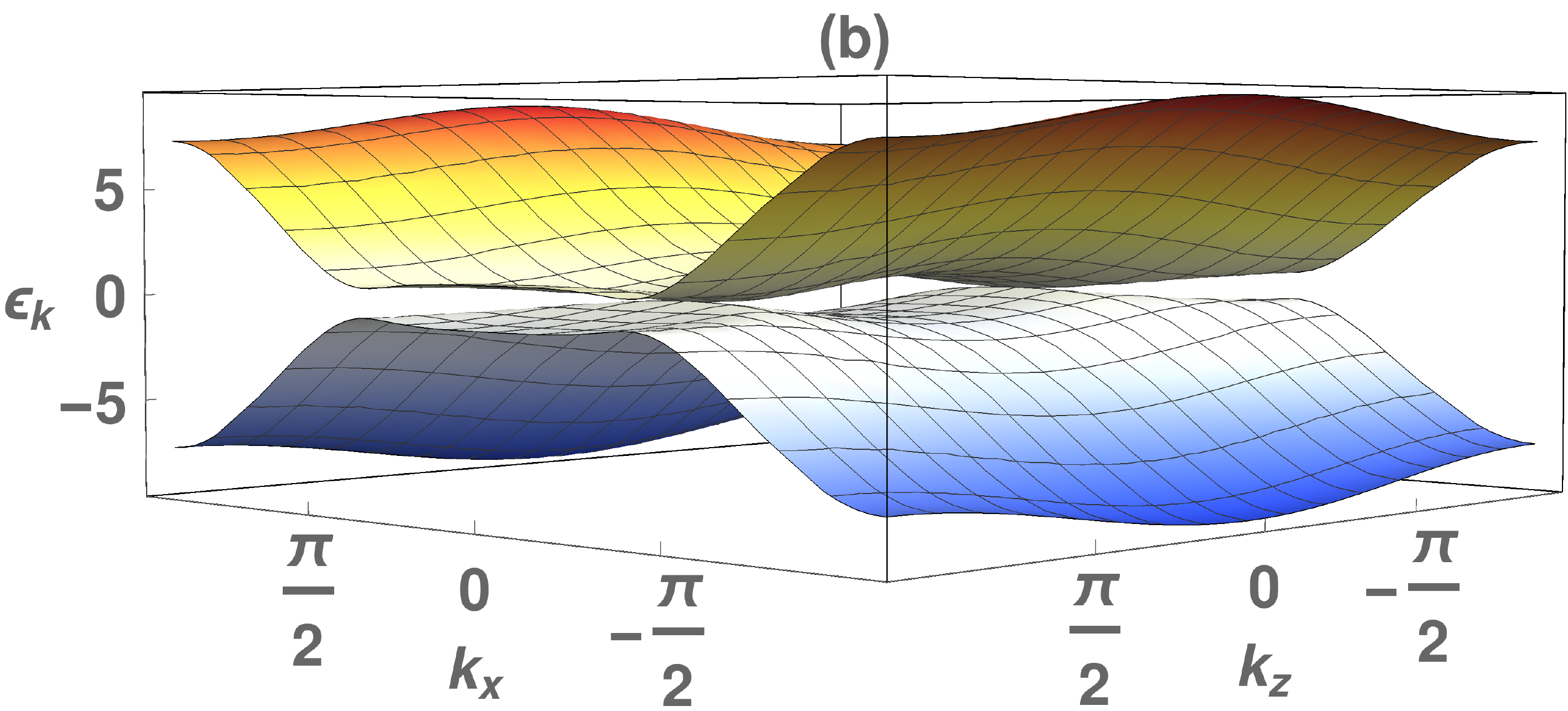,trim=0.0in 0.05in 0.0in 0.05in,clip=true, width=55mm, height=40mm}\vspace{0em}
\epsfig{file=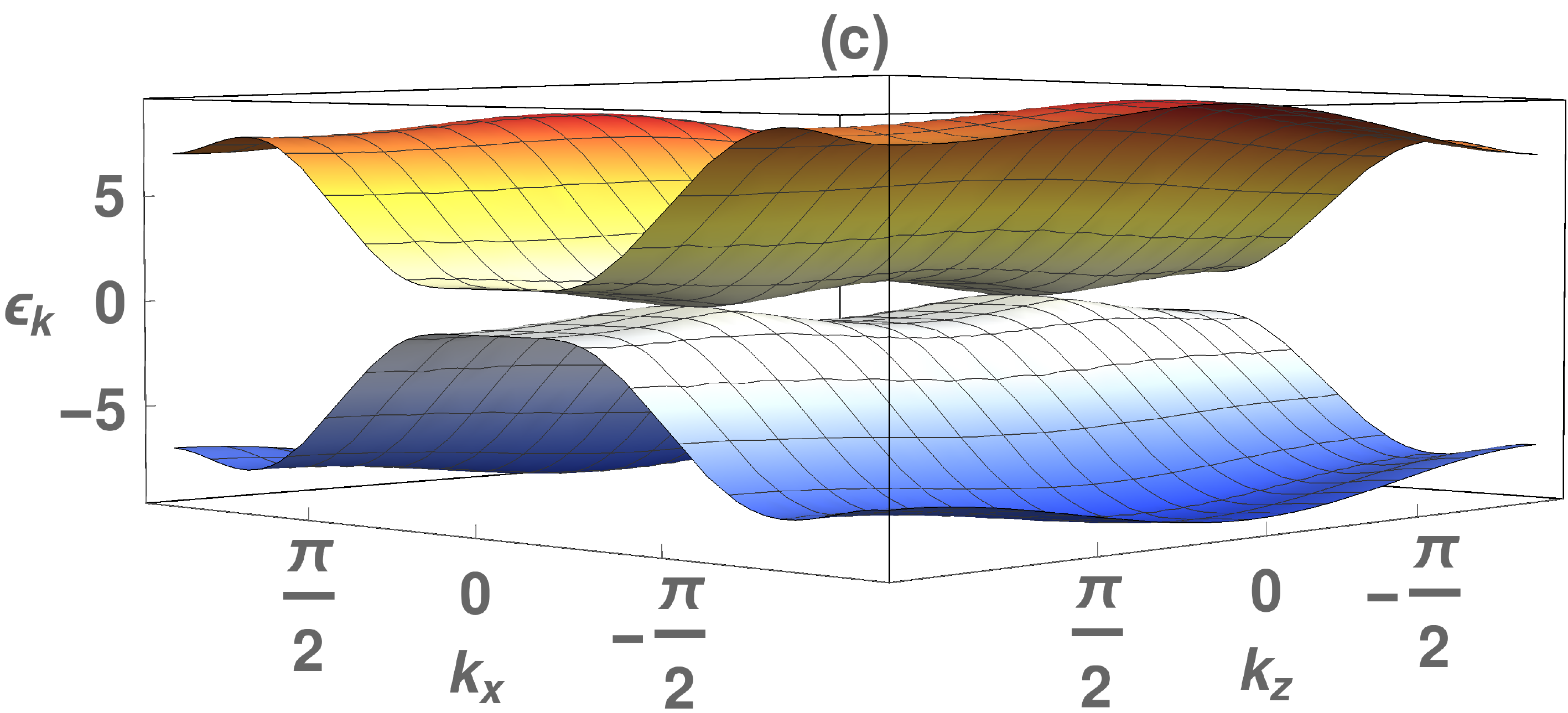,trim=0.0in 0.05in 0.0in 0.05in,clip=true, width=55mm, height=40mm}\vspace{0em}
\caption{(Color online) \textcolor{black}{ The  3D band dispersions of the lattice model of multi-Weyl fermions as presented in Sec.~\ref{s2ss2},
 for (a) $n=1$, (b) $n=2$ and (c) $n=3$ respectively. The energy $\epsilon_{\mathbf{k}}$ is measured in units of eV. The chemical potential is set at zero energy and the lattice constant is taken $a=1$. 
The Weyl nodes are at ($0,0,k_{0}$) and ($0,0,-k_{0}$). We consider $m_z=0.0, t=t_0=t_z=1.0$, and $k_y=0.0$.
The anisotropic nature is vividly noticed for $n>1$. }}
\label{lat_dis}
\end{center}
\end{figure*}

\begin{figure*}[t]
\begin{center}
\epsfig{file=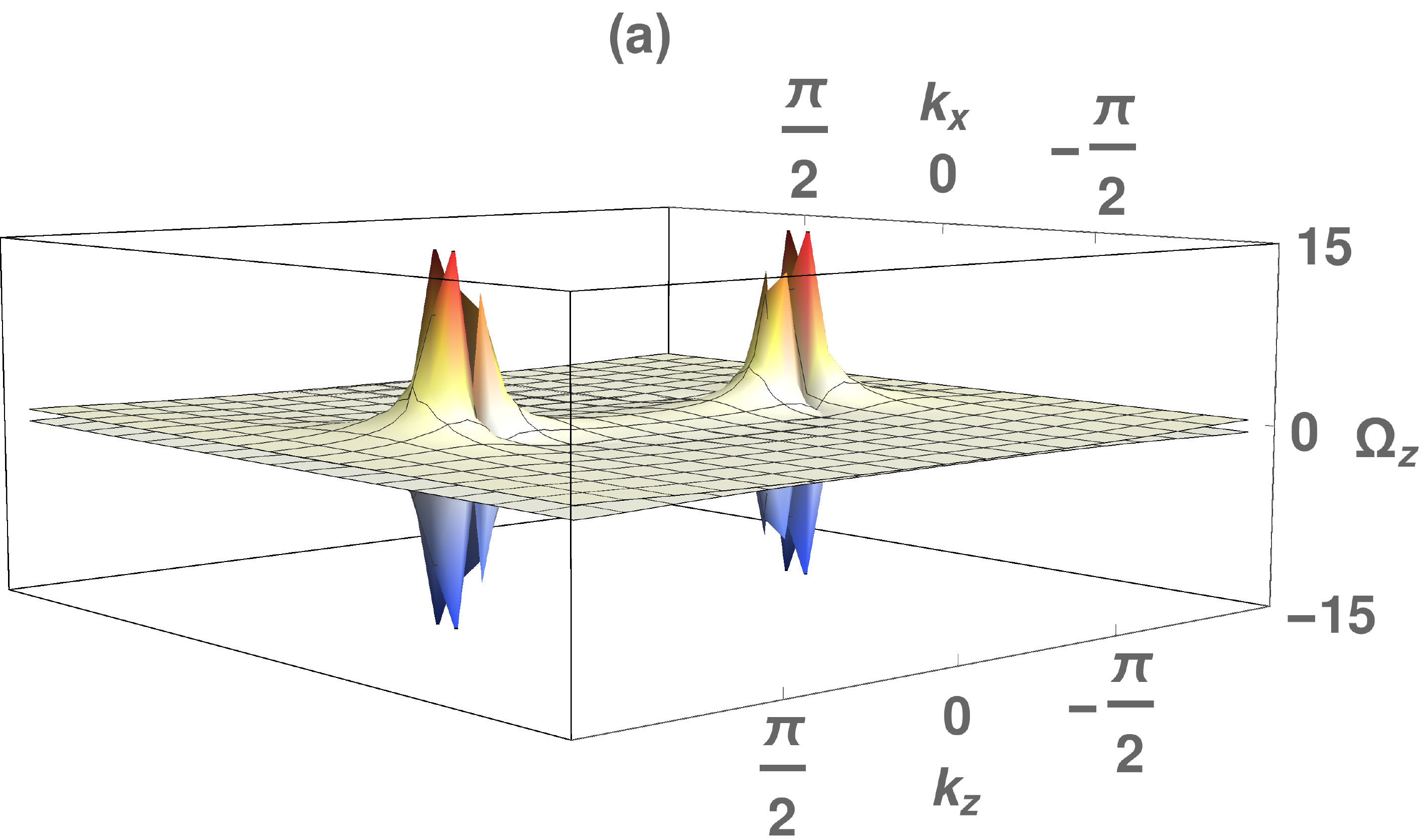,trim=0.0in 0.05in 0.0in 0.05in,clip=true, width=75mm, height=55mm}\vspace{0em}
\epsfig{file=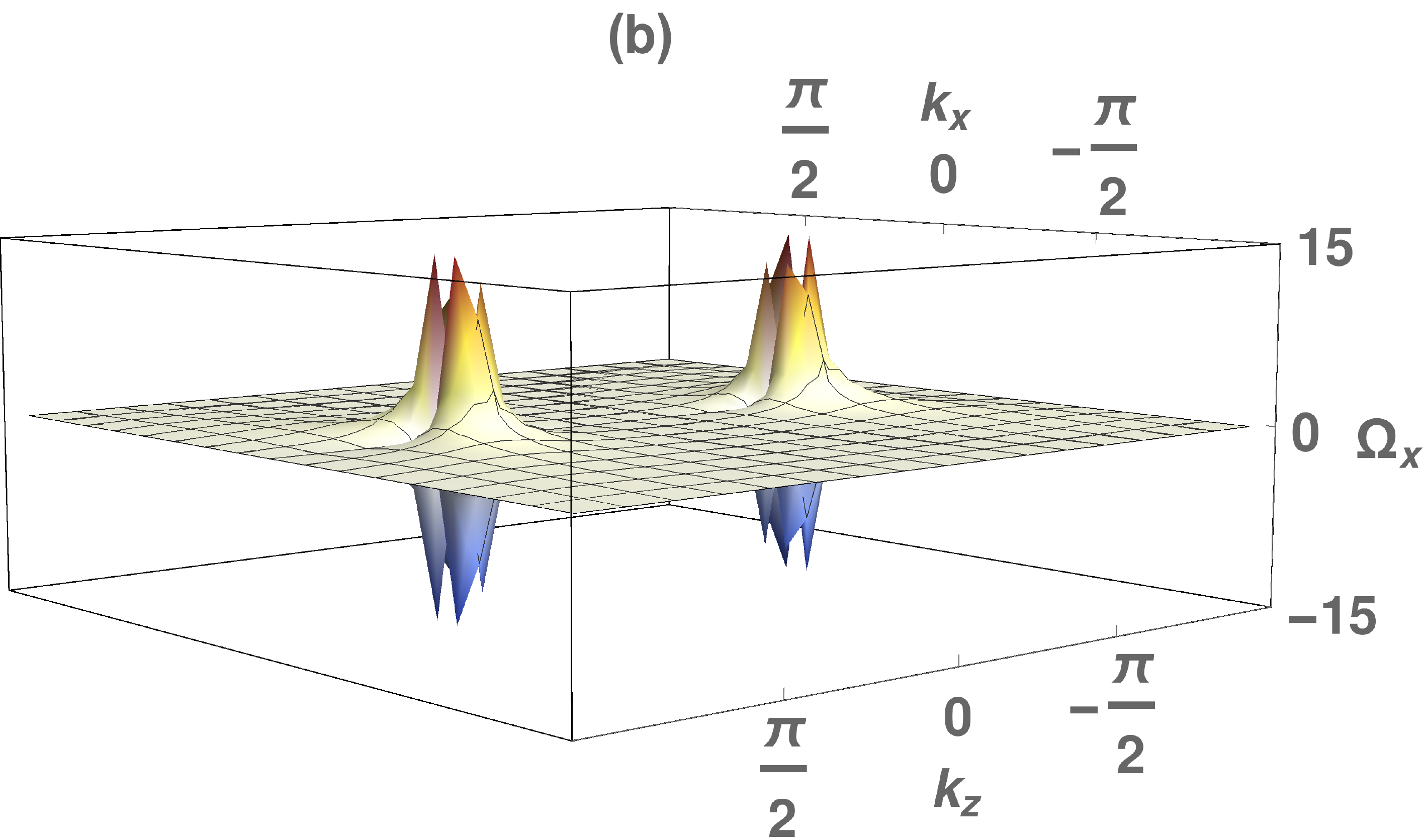,trim=0.0in 0.05in 0.0in 0.05in,clip=true, width=75mm, height=55mm}\vspace{0em}\\
\epsfig{file=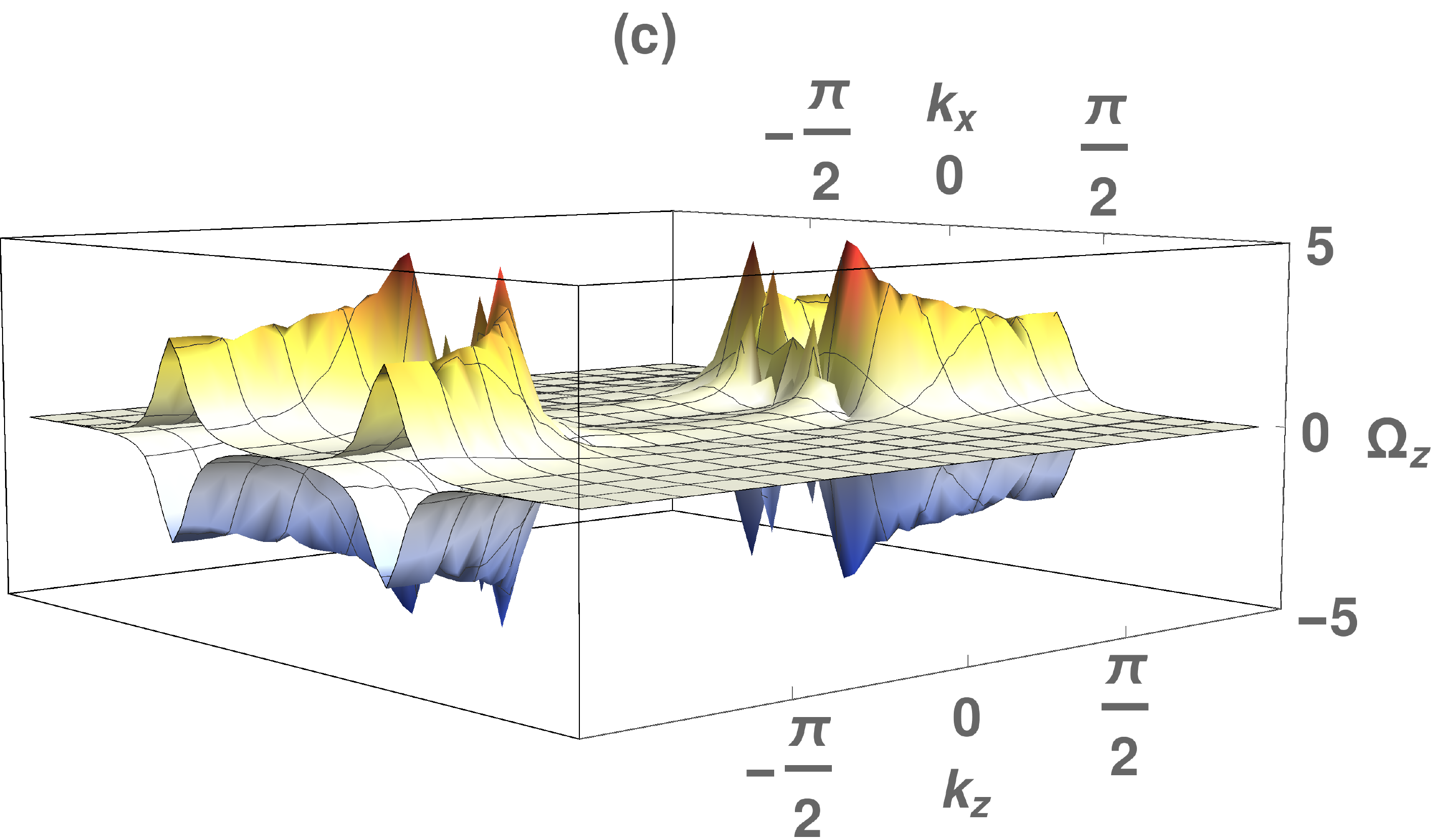,trim=0.0in 0.05in 0.0in 0.05in,clip=true, width=75mm, height=55mm}\vspace{0em}
\epsfig{file=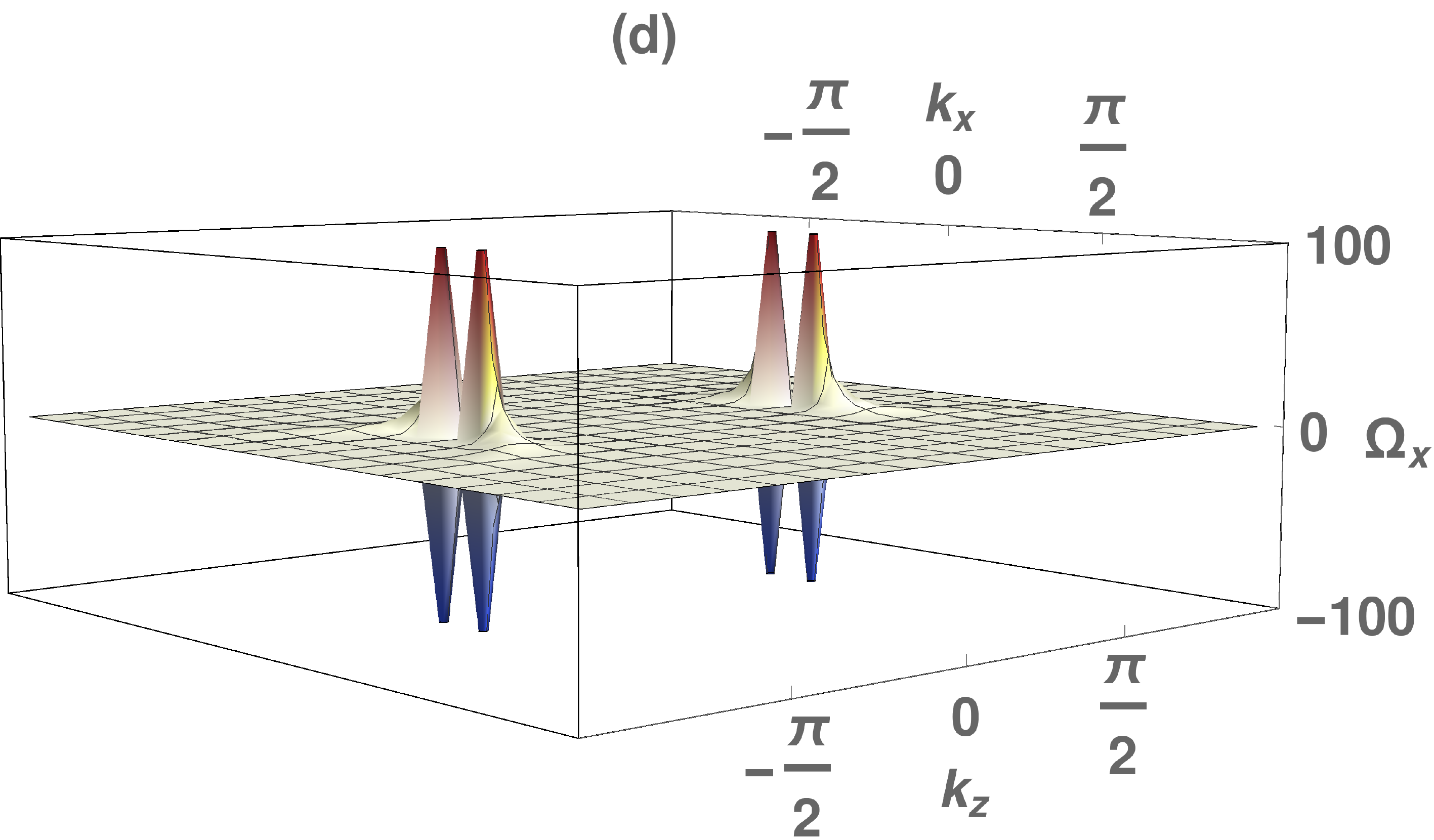,trim=0.0in 0.05in 0.0in 0.05in,clip=true, width=75mm, height=55mm}\vspace{0em}\\
\epsfig{file=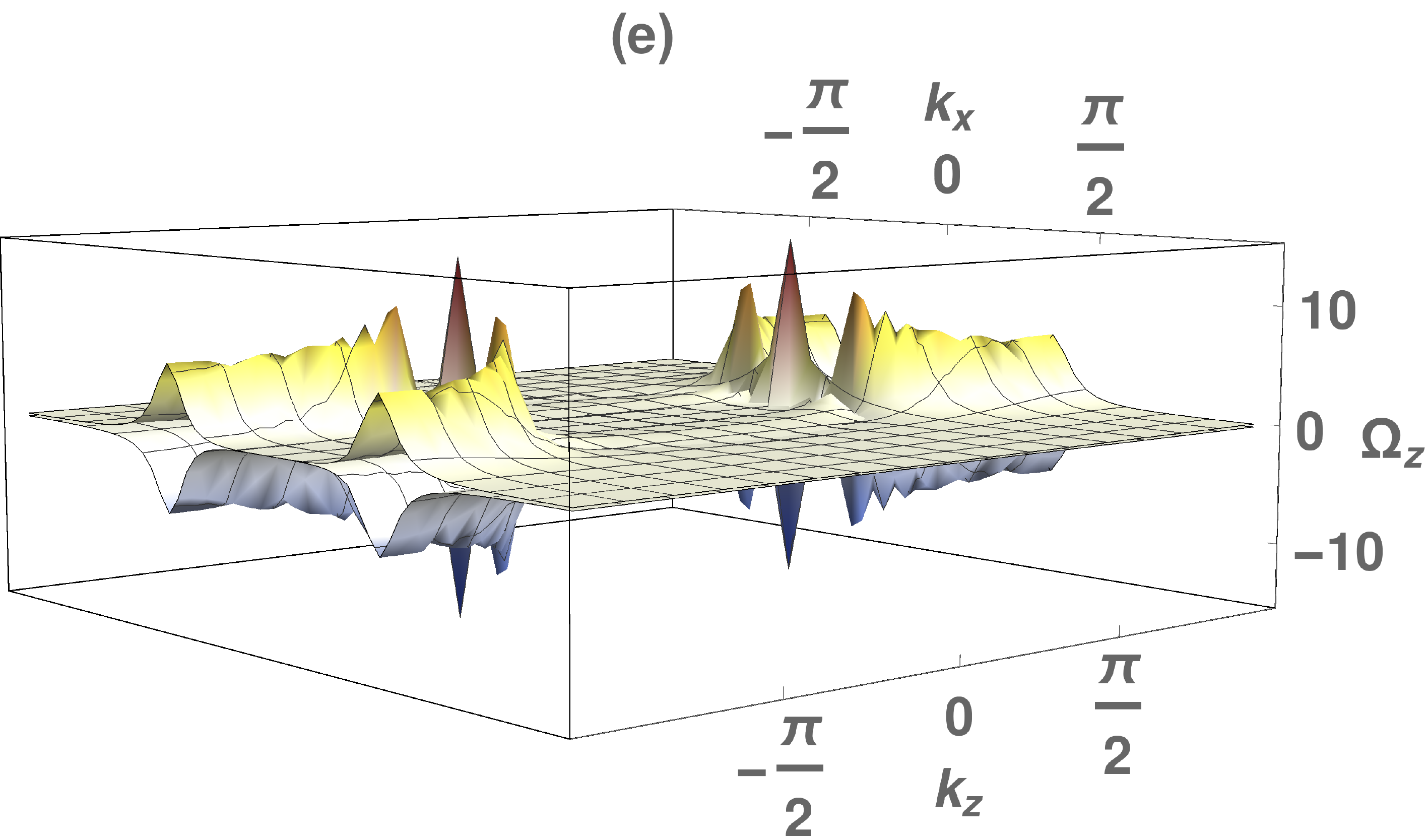,trim=0.0in 0.05in 0.0in 0.05in,clip=true, width=75mm, height=55mm}\vspace{0em}
\epsfig{file=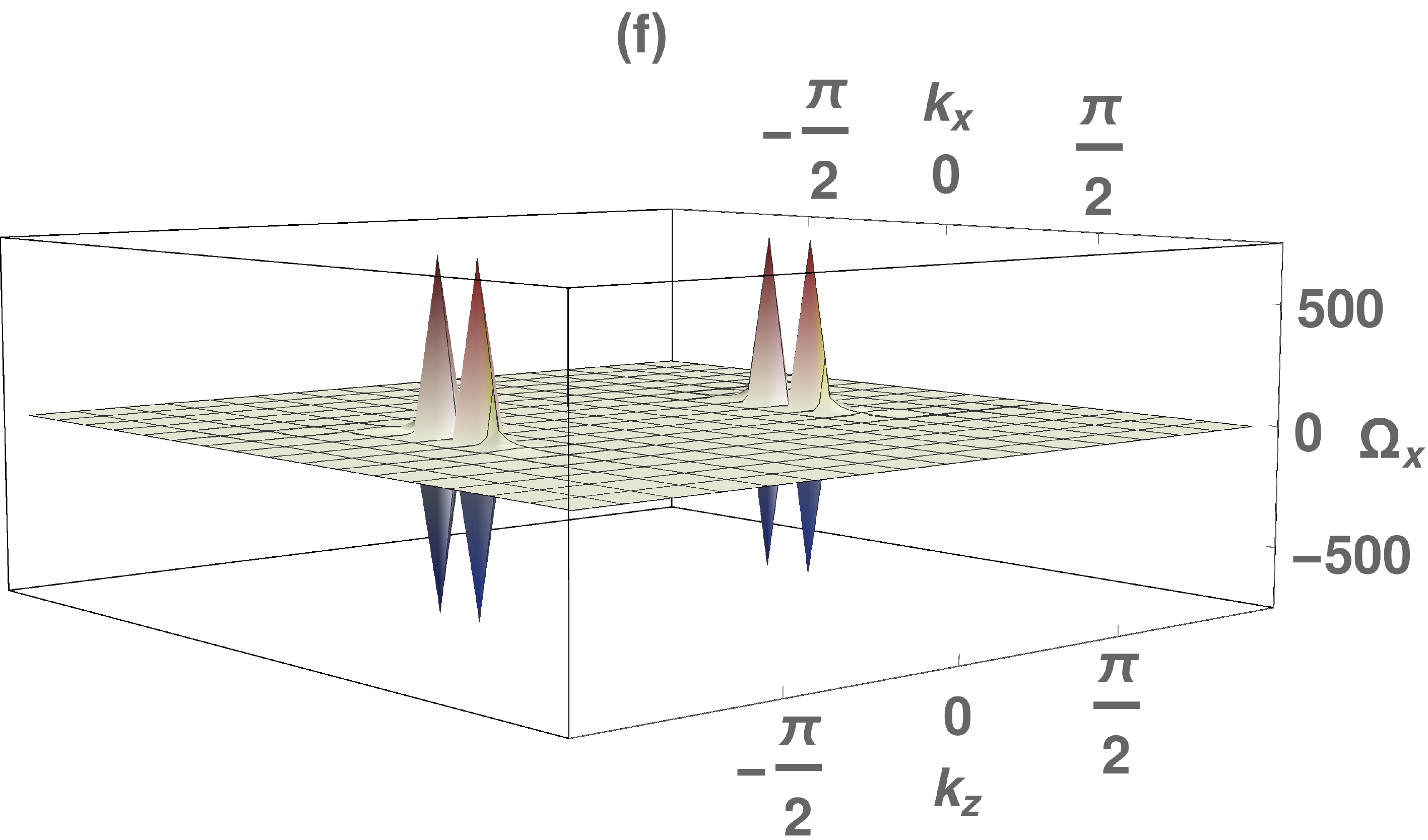,trim=0.0in 0.05in 0.0in 0.05in,clip=true, width=75mm, height=55mm}\vspace{0em}
\caption{(Color online) \textcolor{black}{The Berry curvature components $\Omega_z$ and $\Omega_x$, obtained using Eq.~\ref{bc_lattice}, considering the lattice model in Sec.~\ref{s2ss2} are shown for $n=1$ in (a), (b), 
for $n=2$ in (c), (d) and for $n=3$ in (e), (f), respectively.
We consider here $m_z=0.0, t=t_0=t_z=1.0$ and $k_y=0.025$ for numerical calculation. 
The anisotropic nature between $\Omega_x$ and $\Omega_z$ components can be clearly 
visible for $n=2$ and $n=3$.}}
\label{berry_continuum}
\end{center}
\end{figure*}


\subsection{Lattice Hamiltonian}
\label{s2ss2}
We now discuss a prototype lattice model for \textcolor{black}{type-I} m-WSM that breaks TRS but remains
invariant under inversion.
The corresponding lattice model can be written as~\cite{Roy_2017}

\be
H=\mathbf{N}_{\mathbf{k}}\cdot \boldsymbol{\sigma}.
\ee
For the single WSM with $n=1$, \textcolor{black}{the momentum-dependent form factors $\mathbf{N}_{\mathbf{k}}$ (setting the lattice constant $a = 1$)} takes the form 
$N_x=t \sin k_x$, $N_y=t \sin k_y$ and $N_z= t_z \cos k_z -m_z +t_0(2-\cos k_x -\cos k_y)$.
In this model, the Weyl nodes are located at ${\mathbf k}=(0,0,\pm k_0)$ with 
\be
\cos (k_0)=\frac{t_0}{t_z}\bigl[ \frac{m_z}{t_0}+\cos k_x +\cos k_y -2\bigr].
\label{eq_lm1}
\ee

On the other hand, in the case of a double WSM ($n=2$), the form of $N_{k}$ becomes 
$N_x=t(\cos k_x-\cos k_y)$, $N_y=t\sin k_x \sin k_y$ and $N_z=t_z \cos k_z -m_z + t_0(6+\cos 2k_x
+\cos 2k_y-4 \cos k_x -4 \cos k_y)$. The lattice model of double WSM contains two Weyl nodes at $(0,0,\pm k_0)$ with 
\be
\cos (k_0)=\frac{t_0}{t_z}\bigl[ \frac{m_z}{t_0}-(6+\cos 2k_x +\cos 2k_y-4\cos k_x -4\cos k_y)\bigr].
\label{eq_lm2}
\ee

Similarly, for a triple-WSM with the topological charge $n=3$, one should replace $N_{k}$ by 
 $N_x=t\sin k_x[1-\cos k_x-3(1-\cos k_y)]$, $N_y=-t\sin k_y [1-\cos k_y -3(1-\cos k_x)]$ and $N_z=
 t_z \cos k_z -m_z + t_0(6+\cos 2k_x
+\cos 2k_y-4 \cos k_x -4 \cos k_y)$. Here, the Weyl points appear at ${\mathbf k}=(0,0,\pm k_0)$ 
with $k_0$ followed by the Eq.~(\ref{eq_lm2}). 
The energy dispersions of single, double and triple WSMs along various high-symmetry directions are shown in Fig.~\ref{lat_dis}. 
\textcolor{black}{We note that one can obtain low-energy Hamiltonian (\ref{eq_multi1}) from the above-mentioned lattice Hamiltonian by 
suitably expanding around the gap closing momentum $k_0$. The energy dispersion and the Berry curvature, obtained from lattice model, are shown in Fig.~\ref{lat_dis} and Fig.~\ref{berry_continuum}, respectively.} 
 

\section{Semiclassical formalism for calculating transport coefficients}
\label{Boltzmann}

It has been shown that in the presence of electric field and magnetic field, transport 
properties get substantially modified due to the presence of non-trivial Berry curvature 
which acts as a fictitious magnetic field in the momentum space~\cite{Xiao_2010}. \textcolor{black}{In this section, 
using semiclassical Boltzmann transport theory,
we present general expression of some specific transport properties, namely, LMC,
PHC and TECs that could generally be observed in all Dirac and Weyl
semimetals. In this regime, we consider $T \ll \sqrt{B} \ll \mu$, where $\mu$ is the chemical potential, measured 
from the band-touching point and ignore the Landau quantization of the energy levels.}

In the presence of external perturbative fields (for example, electric field $\mathbf{E}$ and
temperature gradient $\mathbf{\nabla T}$), the charge current $\mathbf{J}$ and thermal current $\mathbf{Q}$
from linear response theory, can be written as
\begin{equation}
J_{\alpha}=L_{\alpha\beta}^{11}E_{\beta}+L_{\alpha\beta}^{12}(-\nabla_{\beta} T),
\label{e02}
\end{equation}
\begin{equation}
Q_{\alpha}=L_{\alpha\beta}^{21}E_{\beta}+L_{\alpha\beta}^{22}(-\nabla_{\beta} T),
\label{e03}
\end{equation}
where $\alpha$ and $\beta$ are spatial indices running over $x$, $y$, $z$. Here, $L_{\alpha\beta}^{11}$
and $L_{\alpha\beta}^{12}$ define the charge conductivity tensor and thermo-electric tensor respectively.
The tensors $L_{\alpha\beta}^{12}$ and $L_{\alpha\beta}^{21}$ are related to each other by the Onsager's
relation : $L_{\alpha\beta}^{21}$=T $L_{\alpha\beta}^{12}$.
In the low temperature regime, the transport
coefficients obey the Mott relation~\cite{Mermin} as $L_{\alpha\beta}^{12}=-\frac{\pi^2}{3e}k^2_{B}T \frac{\partial L_{\alpha\beta}^{11}}{\partial \mu}$, \textcolor{black}{where $e$ is the electronic charge and $k_B$ is the Boltzmann constant.}

The Boltzmann transport equation in its' phenomenological form can be written as~\cite{John_2001}
\begin{equation}
\left(\frac{\partial}{\partial t}+\mathbf{\dot{r}}\cdot\mathbf{\nabla_{r}}+\mathbf{\dot{k}}
\cdot\mathbf{\nabla_{k}}\right)f_{\mathbf{k},\mathbf{r},t}=I_{coll} \{f_{\mathbf{k},\mathbf{r},t}\},
\label{eq_BZ}
\end{equation}
where the right side $I_{coll} \{f_{\mathbf{k},\mathbf{r},t}\}$ is the collision integral which
incorporates the effects of electron correlations and impurity scattering. We are interested in computing
the electron distribution function which is given by $f_{\mathbf{k},\mathbf{r},t}$. Under the relaxation time
approximation  with the parameter $\tau$ that quantifies 
the average time between two successive collisions, \textcolor{black}{the
steady-state Boltzmann equation can be written as}
\begin{equation}
(\mathbf{\dot{r}}\cdot\mathbf{\nabla_{r}}+\mathbf{\dot{k}}\cdot\mathbf{\nabla_{k}})f_{\mathbf{k}}=\frac{f_{0}-f_{\mathbf{k}}}{\tau(\mathbf{k})},
\label{eq_BZf}
\end{equation}
where $f_{0}$ is the equilibrium Fermi-Dirac distribution function.
\textcolor{black}{In this work, we ignore the  momentum dependence of $\tau$ for simplifying the calculations and assume it to be a constant
 \cite{Son:2013, Kim:2014, Fiete_2014,Sharma:2016}.} 
Now we shall revisit the semiclassical equations of motion for an electron in presence of Berry curvature~\cite{Son_2012,Duval_2006}
\be
\mathbf{\dot{r}}=D(\mathbf{B,\Omega_{k}})[\mathbf{v_{k}}+\frac{e}{\hbar}(\mathbf{E}\times
\mathbf{\Omega_{k}})+\frac{e}{\hbar}(\mathbf{v_{k}}\cdot\mathbf{\Omega_{k}})\mathbf{B}],
\label{eq_motion1}\\
\ee
\be
\hbar\mathbf{\dot{k}}=D(\mathbf{B,\Omega_{k}})[e\mathbf{E}+\frac{e}{\hbar}(\mathbf{v_{k}}
\times \mathbf{B})+\frac{e^{2}}{\hbar}(\mathbf{E}\cdot\mathbf{B})\mathbf{\Omega_{k}}].
\label{eq_motion2}
\ee
Here, $D(\mathbf{B,\Omega_{k}})=(1+\frac{e}{\hbar}(\mathbf{B}.\mathbf{\Omega_{k}}))^{-1}$ is the phase space
factor as the Berry curvature  $\mathbf{\Omega_{k}}$ modifies the phase space volume element 
$dkdx \rightarrow D(\mathbf{B,\Omega_{k}})dkdx$~\cite{Duval_2006}.
Hereafter, we denote $D(\mathbf{B,\Omega_{k}})$ by $D$. The  term $(\mathbf{E}\times\mathbf{\Omega_{k}})$ represents
the anomalous velocity  perpendicular to the applied electric field. \textcolor{black}{On the other hand, the third term of Eq.~\ref{eq_motion1} $(\mathbf{v_{k}}\cdot\mathbf{\Omega_{k}})\mathbf{B}$ represents the chiral magnetic effect (CME). 
This leads to interesting signature of transport phenomena in Weyl semimetals
and can appears for $\mathbf{E}=0$ (\textcolor{black} {i.e., $\mathbf{E}\cdot\mathbf{B} =0$)}~\cite{Son_2012,Franz_2013,Yin_2012,Chen_2013,Kenji_2008}}.  Moreover, the term \textcolor{black}{ $(\mathbf{E}\cdot\mathbf{B}) \ne 0$} is responsible for chiral anomaly 
which arises in axion-electrodynamics of WSM. 

\subsection{Setup 1: Longitudinal Magneto-Conductivity and Planar Hall Conductivity}
\label{PH}
\begin{figure}[htb]
\begin{center}
\epsfig{file=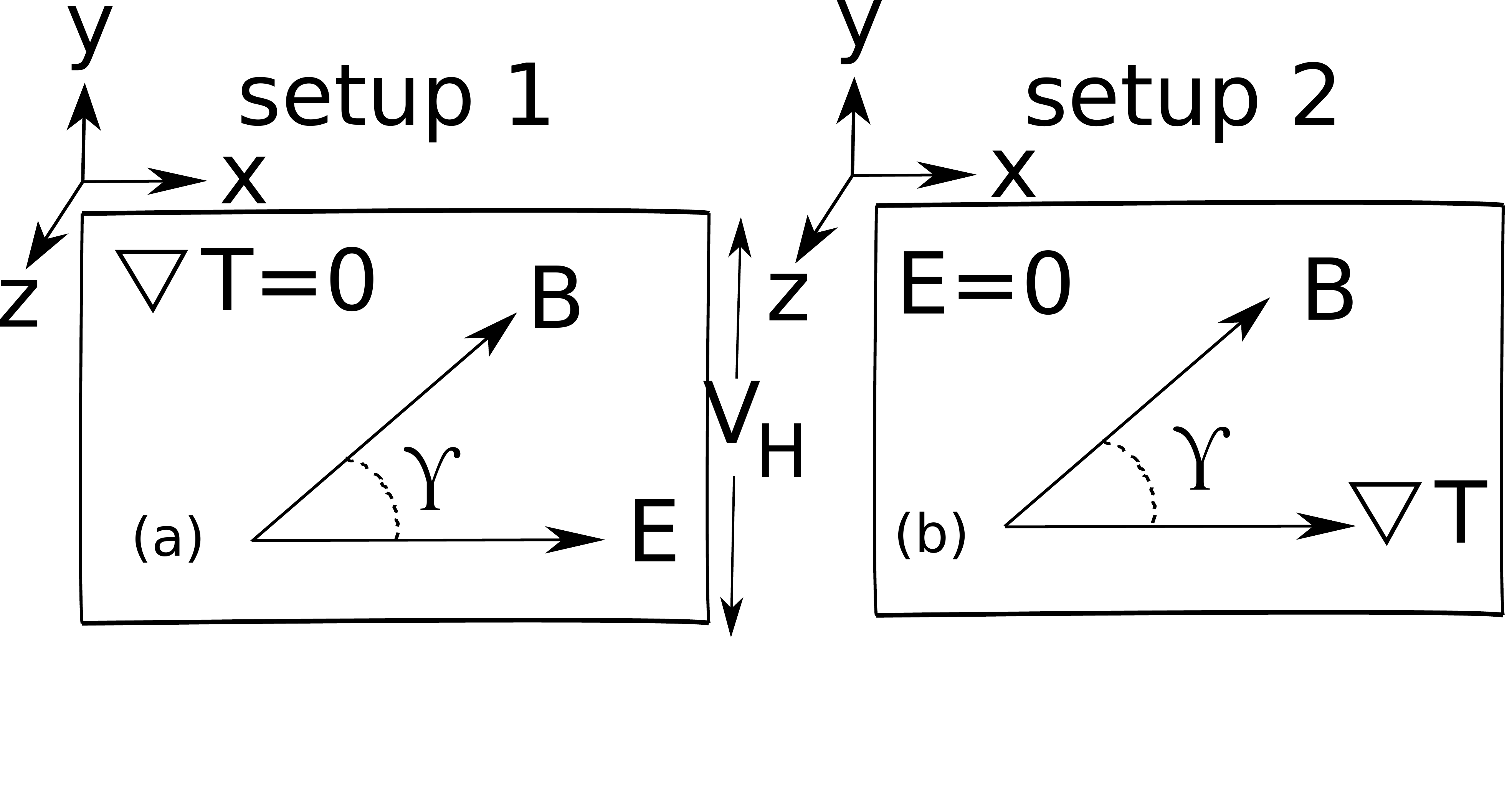,trim=0.0in 0.05in 0.0in 0.05in,clip=true, width=83mm}\vspace{0em}
\caption{(Color online) (a): The geometrical configuration of electric field $\mathbf{E}$ and magnetic field $\mathbf{B}$
is depicted in the planar Hall setup (setup 1). Here, $V_{H}$ is the induced Hall voltage.
(b): The geometrical configuration of temperature gradient $\mathbf{\nabla T}$ and magnetic field $\mathbf{B}$ is depicted in 
planar thermo-electric setup (setup 2).}
\label{setup}
\end{center}
\end{figure}

\textcolor{black}{The PHE is defined through an induction of in-plane transverse voltage when the
co-planar electric and magnetic fields are not perfectly aligned with each other. 
In order to get 
the general expression for PHC and LMC, we consider that the electric field is applied along the $x-$axis and the magnetic field is rotated in $x-y$ plane at a finite angle
$\gamma$ from the $x-$axis,
i.e. $\mathbf{B}=B\cos\gamma \hat{x}+B\sin\gamma\hat{y}$, $\mathbf{E}=E\hat{x}$. The corresponding 
setup 1 is shown in Fig.~\ref{setup}(a)}.

Plugging the equations of motion described in Eq.~(\ref{eq_motion1}) and Eq.~(\ref{eq_motion2}) into
the Boltzmann equation, the general expression of the PHC $\sigma_{yx}$ and LMC $\sigma_{xx}$ in the above configuration can be written
as~\cite{Fiete_2014,Sharma:2016,Nandy_2017, Nandy_2018}
\ba
\sigma_{yx}& \approx e^{2}\int\frac{d^{3}k}{(2\pi)^{3}}D\tau\left(-\frac{\partial f_{0}}{\partial \epsilon}\right) 
[(v_{y}+\frac{eB\sin \gamma}{\hbar}(\mathbf{v_{k}}\cdot\mathbf{\Omega_{k}})) \nonumber \\
&(v_{x}+\frac{eB\cos \gamma}{\hbar}(\mathbf{v_{k}}\cdot\mathbf{\Omega_{k}}))]={L^{11}_{yx}}
\label{eq_ehc}
\ea
and 
\begin{eqnarray}
\sigma_{xx}&& \approx e^{2}\int\frac{d^{3}k}{(2\pi)^{3}}\tau [D({v_{x}}+
\frac{eB\cos \gamma}{\hbar}(\mathbf{v_{k}}\cdot\mathbf{\Omega_{k}}))^{2} 
]\left(-\frac{\partial f_{0}}{\partial \epsilon}\right) \nonumber \\
&&={L^{11}_{xx}}. 
\label{eq_lmc}
\end{eqnarray}

\textcolor{black}{We would like to point out that we use $\approx$ sign in Eq.~(\ref{eq_ehc}) and Eq.~(\ref{eq_lmc})
as we ignore the contribution from the correction factors arising 
due to the presence of external magnetic field. In particular, for the semiclassical regime, 
it is sufficient to retain only the leading order terms in the distribution function $f_k$ as the {contribution from the}
 correction factors are several order of magnitude smaller
than the leading order terms \cite{Nandy_2017} 
(see Appendix \ref{formula_PHC}). The important point to note here is that $B\cos \gamma$ factor associated with 
$\mathbf{v_{k}}\cdot\mathbf{\Omega_{k}}$ in Eq.~(\ref{eq_ehc}) and Eq.~(\ref{eq_lmc})
bears the signature of chiral anomaly (${\mathbf E}\cdot{\mathbf B}$) which we shall investigate below in detail.}

\subsection{Setup 2: Thermo-Electric Coefficient and Planar Nernst Coefficient}
\label{Thermo}

In order to compute the planar TECs, we apply the temperature gradient $\nabla T$ along the $x$-axis and the magnetic field  is rotated in the $x-y$ plane in 
the absence of electric field i.e. $\mathbf{B}=B\cos\gamma\hat{x}+B\sin\gamma\hat{y}$, $\mathbf{\nabla T}=\nabla 
 T\hat{x}$, $\mathbf{E}=0$. \textcolor{black}{The planar thermo-electric setup (setup 2) is shown in Fig.~\ref{setup}(b).}
Using the equations of motion and semiclassical Boltzmann equation, one can write the TTEC $\alpha_{yx}$ 
and LTEC $\alpha_{xx}$ in this setup as~\cite{Fiete_2014,Sharma:2016}
\ba
\alpha_{yx}& \approx e\int\frac{d^{3}k}{(2\pi)^{3}}\tau \frac{(\epsilon-\mu)}{T} 
\left(-\frac{\partial f_{0}}{\partial \epsilon}\right) [(v_{y}+\frac{eB\sin \gamma}
{\hbar}(\mathbf{v_{k}}\cdot\mathbf{\Omega_{k}})) \nonumber \\
&(v_{x}+\frac{eB\cos \gamma}{\hbar}(\mathbf{v_{k}}\cdot\mathbf{\Omega_{k}}))]={L^{12}_{yx}}
\label{eq_pn7}
\ea
and 
\begin{eqnarray}
\alpha_{xx}&& \approx e\int\frac{d^{3}k}{(2\pi)^{3}}\tau \frac{(\mu-\epsilon)}{T} ({v_{x}}+
\frac{eB\cos \gamma}{\hbar}(\mathbf{v_{k}}\cdot\mathbf{\Omega_{k}}))^{2} 
\left(-\frac{\partial f_{0}}{\partial \epsilon}\right)\nonumber \\
&&={L^{12}_{xx}}. 
\label{eq_pn6}
\end{eqnarray}
\textcolor{black}{Similar to LMC and PHC, in the expression of LTEC and TTEC, we ignore the contribution from the correction factors due to presence of external magnetic field. While passing by we can 
comment that $\mathbf{v_{k}}\cdot\mathbf{\Omega_{k}}$ is the key ingredient for chiral magnetic effect and 
the associated $B\cos \gamma$ factor in Eq.~(\ref{eq_pn7}) and Eq.~(\ref{eq_pn6}) is coming from ${\mathbf B}\cdot  \nabla {\mathbf T}$.}

\textcolor{black}{Now we will formulate the planar Nernst effect (PNE) which is characterized 
by coplanar $\nabla {\mathbf T}$, ${\mathbf E}$ and ${\mathbf B}$. In setup 2, 
the longitudinal temperature gradient $\nabla_x {\mathbf T}$ produces a transverse electric 
field $E_y$ as a result of the coplanar component of the ${\mathbf B}$ field; this is
known as PNE.
Unlike the conventional and anomalous Nernst effects, 
PNE appears when the ${\nabla} {\mathbf T}$ and the magnetic field $\mathbf{B}$ are not aligned with each
other. Using the charge conductivity tensor $\sigma$ and thermo-electric tensor $\alpha$,
the PNC $\nu$ can be written as~\cite{Sharma:2016}}
\be
\nu=\frac{E_y}{-dT/dx}=\frac{\al_{xy}\sigma_{xx}-\al_{xx}\sigma_{xy}}{\sigma_{xx}^2 + \sigma_{xy}^2}.
\label{nrnst_eq1}
\ee

In general, the generation of a transverse electric field in the presence of a transverse temperature 
gradient refers to the Nernst effect. The conventional Nernst effect appears due to Lorentz force in a system in the presence 
of an external magnetic field ${\mathbf B}$ applied perpendicular to the temperature gradient $\nabla {\mathbf T}$. The anomalous Nernst effect appears only due to the anomalous velocity of the quasiparticle 
generated by the non-trivial Berry curvature \textcolor{black}{in the absence of external magnetic field. Actually, the conventional (anomalous) Nernst effect requires the finite magnetic field (Berry curvature) in a direction perpendicular to the plane of applied $\nabla {\mathbf T}$ and the induced voltage. On the other hand, this setup will generate an in-plane induced voltage normal
to applied in-plane {$\nabla {\mathbf T}$} and the induced electric field, applied $\nabla {\mathbf T}$ and ${\bm B}$ all lie in the same plane.} 
Therefore, one can infer that the PNE is fundamentally different from the conventional as well as anomalous Nernst effects.

\section{Results}
\label{result}
\textcolor{black}{In this section, we study several intriguing transport properties such as 
LMC, PHC, TECs and PNC using the low-energy model as well as the lattice model of m-WSM.
Using the low energy model, we first calculate these transport coefficients analytically within 
the semiclassical regime and after that we verify our analytical findings by considering the TRS
breaking lattice model. }

\subsection{Analytical Results Using Low Energy Model}
\label{Lin_Results1}

\textcolor{black}{In order to study LMC in m-WSMs, we first breakdown
the complete expression of LMC as given in  Eq.~(\ref{eq_lmc}) into three terms,
(1) $\sigma^{(1)}_{xx}$ $\sim v_x^2$, (2) $\sigma^{(2)}_{xx}$  $\sim ({\mathbf \Omega_{\bm k}}.{\mathbf v_{\bm k}})^2$,
and (3) $\sigma^{(3)}_{xx}$ $\sim v_x({\bm \Omega_{\bm k}}.{\mathbf v_{\bm k}})$. From the Eq.~(\ref{eq_lmc}), 
it is clear that the terms (2) and (3) give together CA induced LMC in m-WSMs. 
This is due to the fact that both (2) and (3) contain the prefactor $B \cos \gamma$ originated from the 
CA (${\mathbf E}\cdot {\mathbf B}$).
Since
it is known that $\sigma^{(2)}_{xx}$ is the most dominant contribution to CA induced LMC in single WSMs without tilt \cite{Nandy_2017}, we will now refer $\sigma^{(2)}_{xx}$ term as $\sigma_{xx}(\rm CA)$ for the rest of 
the work.} 
 
We now analytically calculate each term of LMC as well as the total LMC using the
low-energy model of m-WSMs. The detailed calculations are shown in the Appendix \ref{app2}. 
The CA term $\sigma_{xx}^{(2)}$ and the total LMC are given by 
\ba
\sigma_{xx}({\rm CA}) 
 &&= \eta B^2 \cos^2\gamma ~[ n^3 \mu^{-\frac{2}{n}} +  n  T^2 \mu^{-2-2/n} ]+ {\mathcal O}(T^2n^2),\non \\
 \label{eq_xx2t_1}
 \ea
 \ba
\sigma_{xx}
 && =  \eta \Bigg[ B^2  \cos^2 \gamma  [n^3\mu^{-2/n}+(n+n^2) T^2 \mu^{-2-2/n} ]+ n \mu^2 \non \\
 && + n T^2 \Bigg ] + {\mathcal O}(T^2n^2),
 \label{eq_xx2t_11}
 \ea
\textcolor{black}{
where $\eta=\frac{ v \tau e^4 \al^{2/n}_n }
{16\hbar^2 \pi^{3/2}}\frac{\Gamma(2-1/n)}{\Gamma(5/2-1/n)} $}. 

\textcolor{black}{We shall now examine the LMC in detail as a function of $n$, $\mu$, and $T$. 
It is clear from the Eq.~(\ref{eq_xx2t_1}) and Eq.~(\ref{eq_xx2t_11})
that both $\sigma_{xx}(\rm CA)$ and  $\sigma_{xx}$ vary as $n^3 B^2 \cos^2 \gamma$ at zero 
temperature. The first term $\sigma_{xx}^{(1)}$ (containing velocity part only) yields $B$-independent
contribution (generally referred to as the Drude contribution) to LMC and it varies linearly with 
the topological charge. Therefore, the magnitude of LMC increases as $n^3$
for WSMs with higher $n$~\cite{Roy_2018}. We also find that $\sigma_{xx}(\rm CA)$
is the most dominant contribution to LMC in all WSMs ($n=1,2,3$). 
Moreover, the magnitude of chiral anomaly induced LMC decreases slowly with the chemical potential
$\mu$ in double (scales as $ \mu^{-1}$) and triple (scales as $\mu^{-2/3}$) WSMs compared to single WSM 
(scales as $ \mu^{-2}$) whereas the Drude contribution increases with $\mu^2$ in all WSMs. The multi-Weyl nature thus 
enters into the LMC through the monopole charge. 
On the other hand, at finite temperature, both CA contribution and Drude contribution 
to LMC follow $T^2$ dependence for all WSMs. Interestingly, the temperature dependent CA
induced LMC is proportional to $(n+n^2)\mu^{-2-2/n}$ while the Drude part becomes 
$\mu$ independent and linearly proportional to $n$.}

To investigate the PHC in m-WSMs, we now break the expression of PHC
(as given in Eq.~(\ref{eq_ehc})) similar to the LMC in following form:
 (1) $\sigma^{(1)}_{yx}$ $\sim v_x v_y$, (2) $\sigma^{(2)}_{yx}$ $\sim
 ({\mathbf  \Omega_{\bm k}}.{\mathbf v_{\bm k}})^2$, 
 (3) $\sigma^{(3)}_{yx}$ $\sim v_x({\mathbf \Omega_{\bm k}}.{\mathbf v_{\bm k}})$  and
 (4) $\sigma^{(4)}_{yx}$ $\sim v_y({\mathbf \Omega_{\bm k}}.{\mathbf v_{\bm k}})$. \textcolor{black}{It is clear that 
 the terms (2) and (4), containing $B \cos \gamma$ factor, yield the CA induced PHC as the current flows in $y$-direction.
  Since, in the case of type-I WSM without tilt, term (2) is the most dominant contribution to PHC,
  we below refer $\sigma^{(2)}_{yx}$ as $\si_{yx}(\rm CA)$ for the rest of the paper.}
Using the low-energy model of m-WSMs, we analytically calculate each term of PHC as well as the total PHC. Please see the Appendix \ref{app2} for the detailed calculations.
Now, the $\sigma_{yx}(\rm CA)$ and the total PHC are given by 
\begin{eqnarray}
 \sigma_{yx}({\rm CA}) 
&=& \eta n^3 B^2\cos \gamma \sin \gamma \nonumber \\
&& \times 
 [\mu^{-2/n} + \frac{K_B^2 T^2\pi^2 (2+n)\mu^{-2-2/n}}{3 n^2}] 
 \label{eq_yx2t}
\end{eqnarray}
\begin{eqnarray} 
\sigma_{yx} =\eta B^2  \cos \gamma\sin \gamma  ~[n^3\mu^{-2/n}+(n+n^2) T^2 \mu^{-2-2/n}] \nonumber \\
\label{eq_yx2t1}
\end{eqnarray}  
\textcolor{black}{From Eq.~(\ref{eq_yx2t}) and Eq.~(\ref{eq_yx2t1}), we find that both $\sigma_{yx}({\rm CA})$ and total PHC $\sigma_{yx}$ show $n^3 B^2 \sin\gamma\cos \gamma$ dependence
at $T=0$ in m-WSMs. Unlike LMC, the $B$-independent Drude contribution is zero in the case of PHC.
Therefore, it is clear that the total contribution of PHC is coming from chiral anomaly in m-WSMs. The magnitude of PHC increases as
$n^3$  at zero temperature. Similar to the case of LMC, the PHC at $T=0$ decreases slowly with doping as we go from single WSM ($n=1$) to triple WSM ($n=3$). 
On the other hand, the temperature 
dependent contribution to PHC varies as $(n+n^2)T^2 \mu^{-2-2/n}$.} 

{ We have also calculated LMC ($\sigma_{zz}$) and PHC ($\sigma_{yz}$) using the low-energy model when the electric 
field is applied in the $\hat{z}$-direction.} 
\textcolor{black}{Comparing $\si_{zz}$ and $\si_{xx}$ side by side, we find that 
the qualitative behavior of both $\si_{zz}$ and $\si_{xx}$ are dictated by CA. Hence, their functional dependence with $B$ and $\gamma$ remain unaltered for all WSMs (i.e. for $n=1,2,3$). 
\textcolor{black}{Interestingly, unlike $\sigma_{xx}$ where the velocity term ($\sigma_{xx}^{(1)}$) is proportional to $n\mu^{2}$ at $T=0$, in the case of $\si_{zz}$ the same term becomes $\propto n^{-1}\mu^{2/n}$.}
This leads to some
quantitative differences between $\si_{zz}$ and $\si_{xx}$ in WSMs with $n>1$. 
This is due to the presence of anisotropy in energy dispersion 
(i.e., $\epsilon_k$ is linear with momentum along $z$ direction whereas becomes quadratic and cubic along $y$ direction for double and 
triple WSMs respectively) as well as Berry curvature in these systems.
\textcolor{black}{ Moreover, we also find quantitative differences between $\si_{yz}$ and $\si_{yx}$ in m-WSMs with $n>1$. One can find that $\sigma_{yz}^{(1)}$ varies as $\mu^{2-2/n}$ while $\sigma_{yx}^{(1)}$ goes as $\mu^{-2/n}$.}
The complete calculations of $\si_{zz}$ and $\si_{yz}$ are presented in Appendix \ref{app2}.
}

Next, we study the thermo-electric responses in m-WSMs using the low-energy model.
{We will follow the same prescription for the term-wise breakdown of LTEC and TTEC.} 
$\al_{xx}^{(1)}$ contains quadratic velocity term $v_x^2$, $\al_{xx}^{(2)}$ contains
$({\mathbf \Omega_{\bm k}}. {\mathbf v_{\bm k}})^2$ coming from chiral magnetic effect, and 
$\al_{xx}^{(3)}$ involves  $v_x({\mathbf \Omega_{\bm k}}.{\mathbf v_{\bm k}})$.
We shall hereafter refer $\al_{xx}^{(2)}$ as $\al_{xx}({\rm CME}) $ due to the fact that 
the dominant contribution in $\al_{xx}$ is coming from the bare CME term  $({\mathbf \Omega_{\bm k}}. {\mathbf v_{\bm k}})^2$.
The CME term $\alpha_{xx}^{(2)}$ and the total LTEC are given by
\be
\al_{xx}({\rm CME}) 
 = \rho n^2  B^2  \cos^2 \gamma T \mu^{-1-2/n} + {\mathcal O}(T^3 n^3)
 \label{eq_xx2t_PN}
 \ee
\be
\alpha_{xx} = \rho T ~[-\mu n +B^2 \mu^{-1-2/n}n^2 \cos^2 \gamma] +  {\mathcal O}(T^3 n^3)
\label{eq_xxt_PN}
\ee
 where $\rho= \frac{v \tau e^3  \alpha^{2/n}_n {\Gamma(2-1/n)} K_B^2
 \pi^{1/2} }{48{\Gamma(5/2-1/n)} \hbar^2} $. The detailed calculations are 
 presented in Appendix \ref{app2}.

The term containing linear power of $n$ in Eq.~(\ref{eq_xxt_PN}) is coming from \textcolor{black}
{the {$\al_{xx}^{(1)}$} consisting only the velocity factors.} It is clear from the above equations that both the $\al_{xx}$(CME) and total $\al_{xx}$ vary as $n^2 B^2 \cos^2 \gamma$. \textcolor{black}{$\al_{xx}$(CME) decreases with $\mu$
in a $n$ dependent manner such that for $n=1$, it falls off more rapidly (as $\mu^{-3}$) than that of for $n=2$ whereas it becomes the most slowly decreasing
function of $\mu$ (as $\mu^{-5/3}$) for $n=3$ among all the WSMs.}
\textcolor{black}{Interestingly, we find that although the magnitude of $\al_{xx}$ enhances with the topological charge similar to electrical conductivities, the scaling of $\al_{xx}$ with topological charge ($\propto n^2$) is different compared to $\sigma$ ($\propto n^3$)}. 
\textcolor{black}{
Moreover, LTEC is a more rapidly decaying function of $\mu$ than LMC. Both of the above 
observations can be understood using Mott relation between
 Eq.~(\ref{eq_xx2t_11}) and Eq.~(\ref{eq_xxt_PN}) in the limit  $T \rightarrow 0$.
\textcolor{black}{Therefore, it is clear that their origins are characteristically 
different.}}

Similar to PHC, one can notice that $\al_{yx}^{(2)} \sim ({\bm \Omega_{\bm k}}. {\bm v_{\bm k}})^2$ bears the 
maximum contribution of TTEC as compared to all the other remaining terms 
$\al_{yx}^{(1)} \sim v_x v_y$
$\al_{yx}^{(3)} \sim v_x({\mathbf \Omega_{\bm k}}.{\mathbf v_{\bm k}})$
and $\al_{yx}^{(4)} \sim v_y({\mathbf \Omega_{\bm k}}.{\mathbf v_{\bm k}})$. 
The CME term $\alpha_{yx}^{(2)}$ and the total TTEC are given by
\be
 \al_{yx}({\rm CME})
= \rho
\cos \gamma \sin \gamma  n^2 B^2  T\pi^2\mu^{-1-2/n},
\label{eq_yx2t_PN}
 \ee  
 \be
\al_{yx} = \rho T B^2 n^2 \cos \gamma \sin \gamma\mu^{-1-2/n} +{\mathcal O}(T^3 n^3).
 \label{eq_yxt_PN}
 \ee
It is clear from the above equations that the total TTEC does not have any $B$-independent 
contribution and is dominated by the CME term. We find that the transverse component {of $\alpha$}
varies similarly (varies as $n^2 B^2$) compared to the longitudinal 
component except the angular part which goes as 
$\sin\gamma\cos \gamma$ {as given in Eq.~(\ref{eq_yxt_PN})}. Using the Mott relation, one can obtain the TTEC (Eq.~\ref{eq_yxt_PN}) from PHC (Eq.~\ref{eq_yx2t1}). \textcolor{black}{We also calculate LTEC ($\al_{zz}$) and TTEC ($\al_{yz}$) when the temperature gradient is along $z$-direction (see Appendix \ref{app2}). We find that there exists quantitative difference between $\al_{xx}$ and $\al_{zz}$ as well as between $\al_{yx}$ and $\al_{yz}$ due to anisotropic energy dispersion of WSMs with $n>1$.} 



We now compute the functional dependence of the PNC as we already 
obtained $\sigma$ and $\alpha$. Using Eq.~(\ref{nrnst_eq1}), the functional form of PNC $\nu$
can be written as
 \be
  \nu = B^2  \cos \gamma \sin \gamma ( f_1(n, T, \mu)+ f_2(n, T, \mu)),
  \label{nrnst_ana}
  \ee
with $f_{1,2}$ being complicated functions of $n, T$ and $\mu$. We present the 
detailed form of $f_1(n, T, \mu)$ and $f_2(n, T, \mu)$ in the Appendix \ref{app2}.
Unlike  $\sigma$ and $\al$, it is clear from the functional form of $\nu$ that the topological
charge dependence is not monotonous in this case. This is due to the fact that both the numerator
and the denominator of  $f_1$ and $f_2$ have non-linear products
 consisting of $\mu$, $n$ and $T$. Hence, one can expect that the behavior of $\nu$ for different
 $n$ would strongly 
 depend on the values of $\mu$ and $T$. \textcolor{black}{The angular dependence of PNC 
 is same as the transverse transport coefficients.} 

\textcolor{black}{In summary, we find that the magnitude of all the transport coefficients such as LMC, PHC, TECs and PNC increases with the topological charge. In particular, the electrical conductivities $\sigma$ (LMC and PHC) enhance with $n^3$ while the thermo-electric coefficients (LTEC and TTEC) increase with $n^2$. In the presence of external magnetic field, the longitudinal transport coefficients 
i.e., both electric $\si$ and thermo-electric $\al$,
show $B^2\cos^2 \gamma$ dependence whereas
the transverse transport coefficients follow $B^2 \sin \gamma \cos \gamma $ in type-I m-WSMs without
tilt. Moreover, unlike single WSM, there exists quantitative difference between longitudinal
transport coefficients (e.g. $\sigma_{xx}$ and $\sigma_{zz}$) in 
the case of double and triple WSMs when an external field is applied along the anisotropic 
direction (e.g. $x$-direction) of the underlying energy dispersion.}

 Interestingly, we find that for single WSM, the magnitude of transport coefficients decreases rapidly (as $\mu^{-2}$ for $\sigma$ at $T=0$ and $\mu^{-3}$
for $\alpha$) with doping compared to double and triple WSMs where the magnitude drops as $\mu^{-2/n}$ at $T=0$ for $\sigma$ and $\mu^{-2/n-1}$ for $\alpha$. All of these $n$-dependent scaling come from the fact that the energy dispersion (Eq.~(\ref{eq_multi2})), 
Berry curvature (Eq.~(\ref{eq_bcl})) and the velocity (Eq.~(\ref{eq_vel})) significantly change with $n$. 
\textcolor{black}{Therefore, by looking at the scaling of transport coefficients with the monopole charge, one can distinguish a double and triple WSMs from a single WSM.}
\textcolor{black}{We would like to note that unlike $\mu$, the functional dependence of different transport coefficients in m-WSMs on temperature $T$ (scales as $T^2$ for LMC and PHC and $T$ for LTEC and TTEC) remains unaltered as compared to the 
single Weyl case. 
Furthermore, it has been shown that in m-WSMs, the number of Fermi arc is given by the monopole
charge $n$ \cite{Roy_2020}. Hence, it is expected that transport properties due to surface
states would get enhanced in WSMs with $n>1$ (double and triple WSMs) compared to single WSM 
as the number of available conducting states increases. Therefore, our results suggest that regardless of 
having similar angular and magnetic field dependencies of different magneto-transport properties 
in all cases ($n=1,2,3$), there is a lot of new physics popping up for double and triple WSMs compared to single WSM.}

We would like to point out that all anomalous (in the absence of magnetic field) thermoelectric coefficients in the limit of zero temperature and small chemical potential are found to be proportional to the
integer topological charge of the Weyl nodes in m-WSMs\cite{Gorbar_2017}. However, our results (i.e. the scaling dependence with $n$) are very different and can not be compared with Ref.~\cite{Gorbar_2017} because CA as well as CME which are the origin for the transport
properties discussed in this work are absent in Ref.\cite{Gorbar_2017}. 
\begin{figure}[htb]
\begin{center}
\epsfig{file=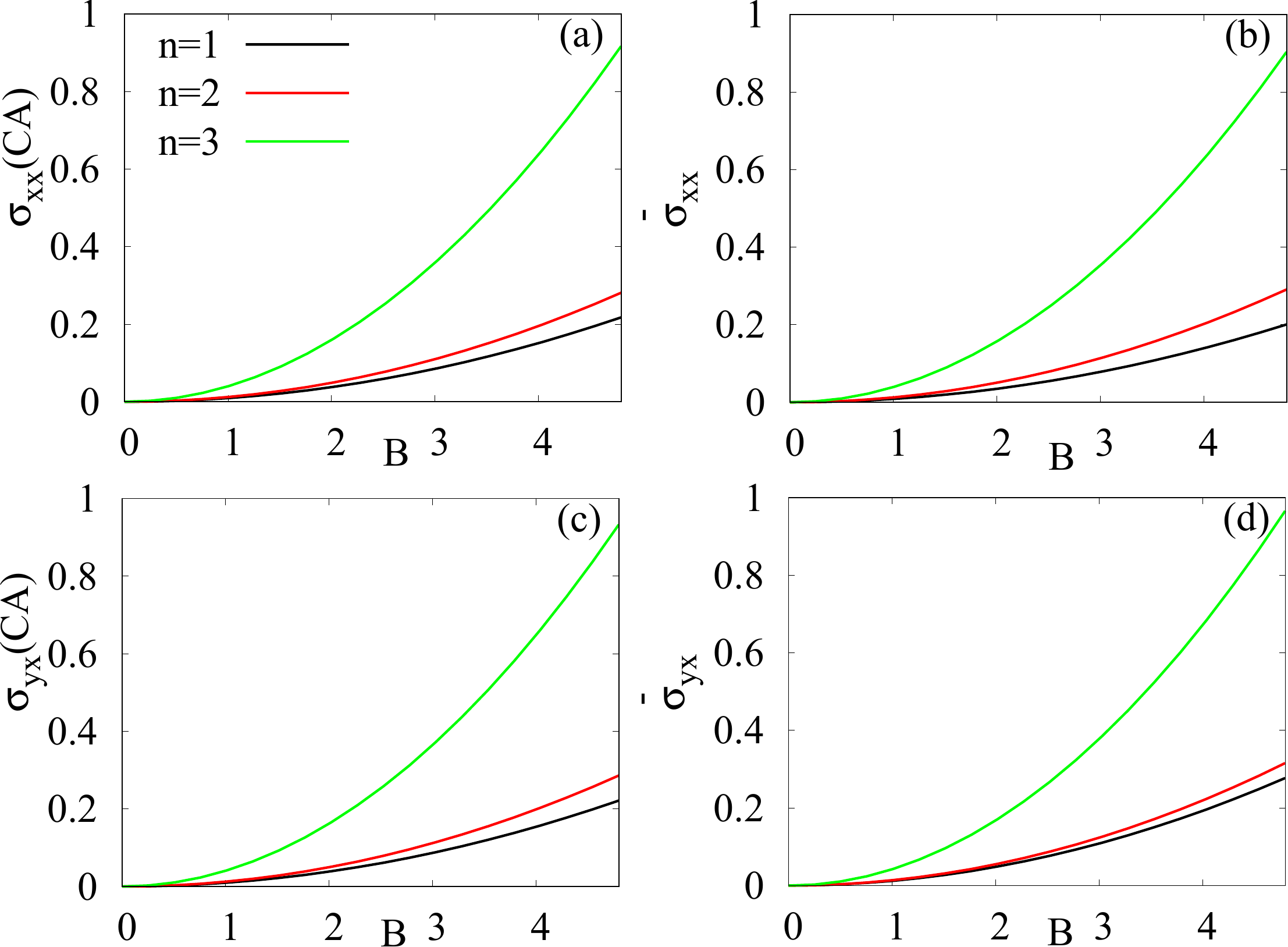,trim=0.0in 0.05in 0.0in 0.05in, width=83mm}\vspace{0em}
\caption{(Color online) \textcolor{black}{(a)-(b) show the behavior of $\sigma_{xx}(\rm{CA})$ and $\bar \sigma_{xx}$ as a function of $B$ for m-WSMs respectively. (c)-(d) depict $\sigma_{yx} (\rm{CA})$ and total PHC $\bar \sigma_{yx}$, calculated using Eq.~\ref{eq_ehc}, as a function of $B$ for m-WSMs. In all the above cases,
the quadratic dependence on $B$ is clearly visible. Both the LMC and the PHC increase with topological charge in a non-linear 
fashion for a given value of $B$. 
The parameters chosen are the following: $\gamma=\pi/3$, $T=10 ~K$, $\mu=0.05 $ eV ( for (a) and (b))
and $\mu=0.07$ eV (for (c) and (d)). }
}
\label{PH_B}
\end{center}
\end{figure}

\begin{figure}[htb]
\begin{center}
\epsfig{file=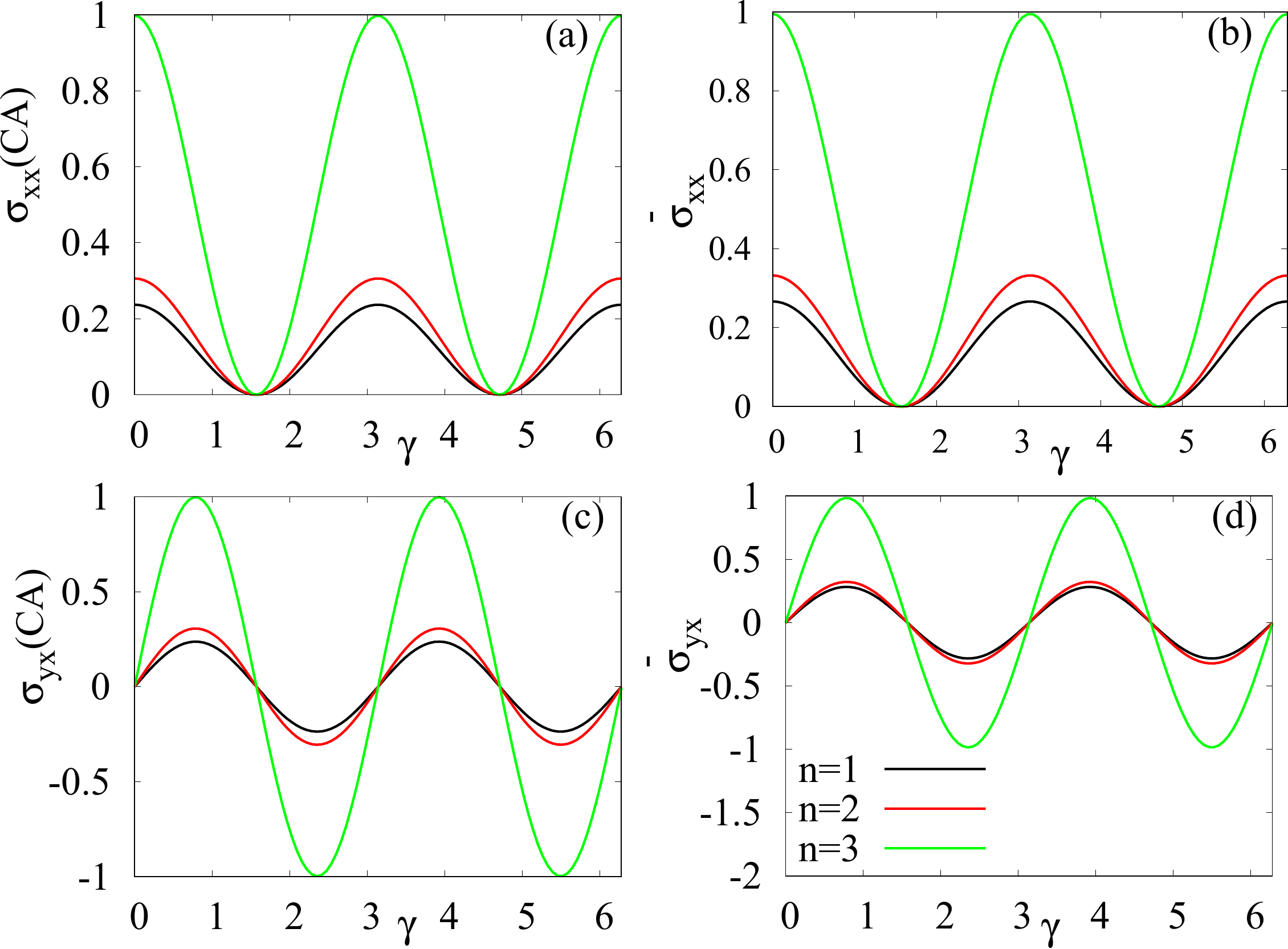,trim=0.0in 0.05in 0.0in 0.05in, width=85mm}\vspace{0em}
\caption{(Color online) \textcolor{black}{(a)-(b) show the behavior of $\sigma_{xx}(\rm{CA})$ and $\bar \sigma_{xx}$ as a function of $\gamma$ at $B=3$ Tesla for m-WSMs respectively. The behavior of $\sigma_{yx} (\rm{CA})$ and total PHC $\bar \sigma_{yx}$ as a function of $\gamma$ at $B=3$ Tesla for m-WSMs are depicted in (c) and (d) respectively. 
It is clear from the figure that LMC and PHC follow $\cos^2 \gamma$ and $\sin \gamma\cos \gamma$
dependence, respectively. Both of them
increase with topological charge in a non-linear fashion for a given value of $\gamma$.
Here, we have taken $\mu=0.05$ eV and $T=10$ K.}
}
\label{PH_th}
\end{center}
\end{figure}

\subsection{Numerical results in Lattice Model}
\label{result_Latt}

In order to discuss transport properties in a physical multi-Weyl system, it is always good to consider a lattice model of Weyl 
fermions with the lattice regularization providing a physical ultra-violet smooth cut-off to the \textcolor{black}{low-energy Dirac spectrum \cite{carbotte16,yago17}.}
\textcolor{black}{Here, we consider tight binding 
lattice model of m-WSMs as discussed in Sec.~\ref{s2ss2} to study electric and thermo-electric responses as a function of $B$ and 
angle $\gamma$.} 

\textcolor{black}{At the outset, we would like to mention 
while showing the variation of
the total contribution of transport quantities with $B$ that we consider $\bar {\si}_{ij}=\si_{ij}(B)-\si_{ij}(B=0)$
and  $\bar {\al}_{ij}=\al_{ij}(B)-\al_{ij}(B=0)$ with $i=j$ for longitudinal and $i\ne j$ for transverse coefficients. 
For the angular variation, in the case of longitudinal component we study the following quantities: $\bar {\si}_{ii}=\si_{ii}(\gamma)-\si_{ii}(\gamma=\pi/2)$
and $\bar {\al}_{ii}=\al_{ii}(\gamma)-\al_{ii}(\gamma=\pi/2)$, while the transverse  components, $i\ne j$, are designated by 
 $\bar {\si}_{ij}=\si_{ij}(\gamma)-\si_{ij}(\gamma=0)$
and $\bar {\al}_{ij}=\al_{ij}(\gamma)-\al_{ij}(\gamma=0)$.}
We now mention that $B$, $T$ and $\mu$ are measured in the units of Tesla, Kelvin and eV respectively.
The conductivities are measured in the units of (Ohm.m)$^{-1}$. \textcolor{black}{We present all the transport 
coefficients in the normalized version for the sake of convenience.}


\subsubsection{Setup 1: Longitudinal Magneto-conductivity and Planar Hall Conductivity}

{\textbf{Magnetic Field:}}
\textcolor{black}{The behavior of total LMC $\bar{\si}_{xx}$ as a
function of magnetic field $B$ for single, double and triple WSMs are shown in Fig.~\ref{PH_B}(b).}
It is clear from the Fig.~\ref{PH_B}(b) that LMC increases quadratically with magnetic field for all cases.
The magnitude of LMC also increases with the topological charge $n$ of the WSMs. 
For detailed investigation, we compute each term of $\si_{xx}$ and find that the leading $B^2$-dependent contribution is coming from the CA
term which is plotted as a function of $B$ in Fig.~\ref{PH_B}(a). 
The variation of total PHC $\sigma_{yx}$ as a function of magnetic field is shown
in Fig.~\ref{PH_B}(d) for $n=1$, $2$ and $3$. One can see that PHC increases in non-linear fashion with $n$ for a particular magnetic field 
and also shows $B^2$ dependence for all WSMs. The {dominating} CA term $\sigma_{yx}({\rm CA})$ is shown in Fig.~\ref{PH_B}(c).
It is evident from this numerical study that the CA is the origin for the appearance of PHC in m-WSMs. 
\textcolor{black}{Therefore, one can infer that our results for LMC and PHC using lattice model qualitatively agree with
the results obtained from low-energy m-WSM model (see Eq.~(\ref{eq_xx2t_11}) and Eq.~(\ref{eq_yx2t1})).} 

{\textbf{Angle}}: 
We will now discuss angular dependence of both LMC and PHC for m-WSMs. 
The  LMC and PHC as a function of the angle $\gamma$ for a particular magnetic field
are depicted in Fig.~\ref{PH_th}(a)-(d). \textcolor{black}{We find that the CA term of LMC ($\sim ({\mathbf \Omega_k}.{\mathbf v_k})^2$)
as well as the total LMC $\bar\si_{xx}$ show $\cos^2 \gamma$ dependence whereas the CA
term of PHC $\sigma_{yx}({\rm CA})$ and total PHC $\bar \sigma_{yx}$ 
exhibit $\sin \gamma \cos \gamma$ dependence irrespective \textcolor{black}{of the value of n.}} The magnitude of both the conductivities
increases with the topological charge associated with the Weyl node. These numerical findings are in full congruence with the analytical results.
We would like to point out that the oscillation  amplitude of $\bar \sigma_{yx}$ for single and double WSMs 
almost coincides with each other whereas the magnitude for triple WSM is much greater than these two.

\begin{figure}[t]
\begin{center}
\epsfig{file=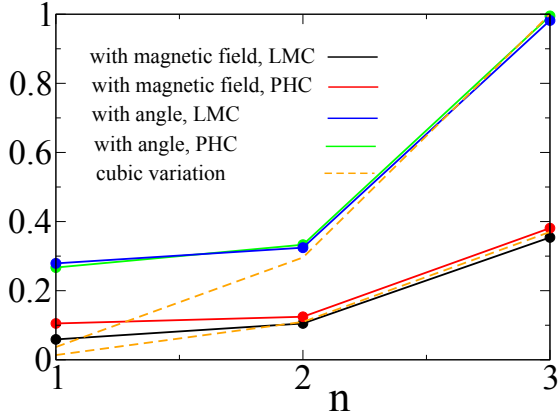,trim=0.0in 0.05in 0.0in 0.05in,clip=true, width=85mm}\vspace{0em}
\caption{(Color online) We plot the variation of LMC and PHC with 
the topological charge $n$. The solid black and blue lines represent the 
data obtained for LMC from magnetic field variation (Fig.~\ref{PH_B}(b))
and angular variation (Fig.~\ref{PH_th}(b)), respectively. The solid red and green lines 
represent the data obtained for PHC from magnetic field variation (Fig.~\ref{PH_B}(d))
and angular variation (Fig.~\ref{PH_th}(d)), respectively.
\textcolor{black}{Both the LMC and the PHC obtained from either magnetic field or 
angular variation show a slow rise as compared to cubic variation (dashed 
yellow line) for $1 \le n \le 2$ while for $2 \le n \le 3$, 
both of them perfectly match with the cubic variation (dashed yellow line).} 
}
\label{PH_n}
\end{center}
\end{figure}

{\textbf{Scaling with ${\mathbf n}$}}: 
The variation of LMC and PHC with topological 
charge is shown in Fig.~\ref{PH_n}. { We consider Fig.~\ref{PH_B}(b), (d) and 
Fig.~\ref{PH_th}(b), (d) to 
investigate the $n$ dependence of LMC and PHC for a given value of $B$ and $\gamma$.}
It is clear from the figure that both LMC and PHC follow cubic variation ($\sim n^3$) for WSMs with $2 \le n \le 3$ while for WSMs with $1 \le n \le 2$,
a deviation from $n^3$ dependence is visible. 
\textcolor{black}{The underlying reason might be related to 
CA term which controls both LMC and PHC maximally for $n>2$. On the other hand, velocity term
might be responsible for this deviation for $n <2$.} 
\textcolor{black}{Therefore, it is clear that due to lattice effects, there is a deviation between the numerical results based on lattice model and analytical results based on low-energy model on monopole charge dependence of LMC and PHC.}




\begin{figure}[ht]
\begin{center}
\includegraphics[width=8.4cm]{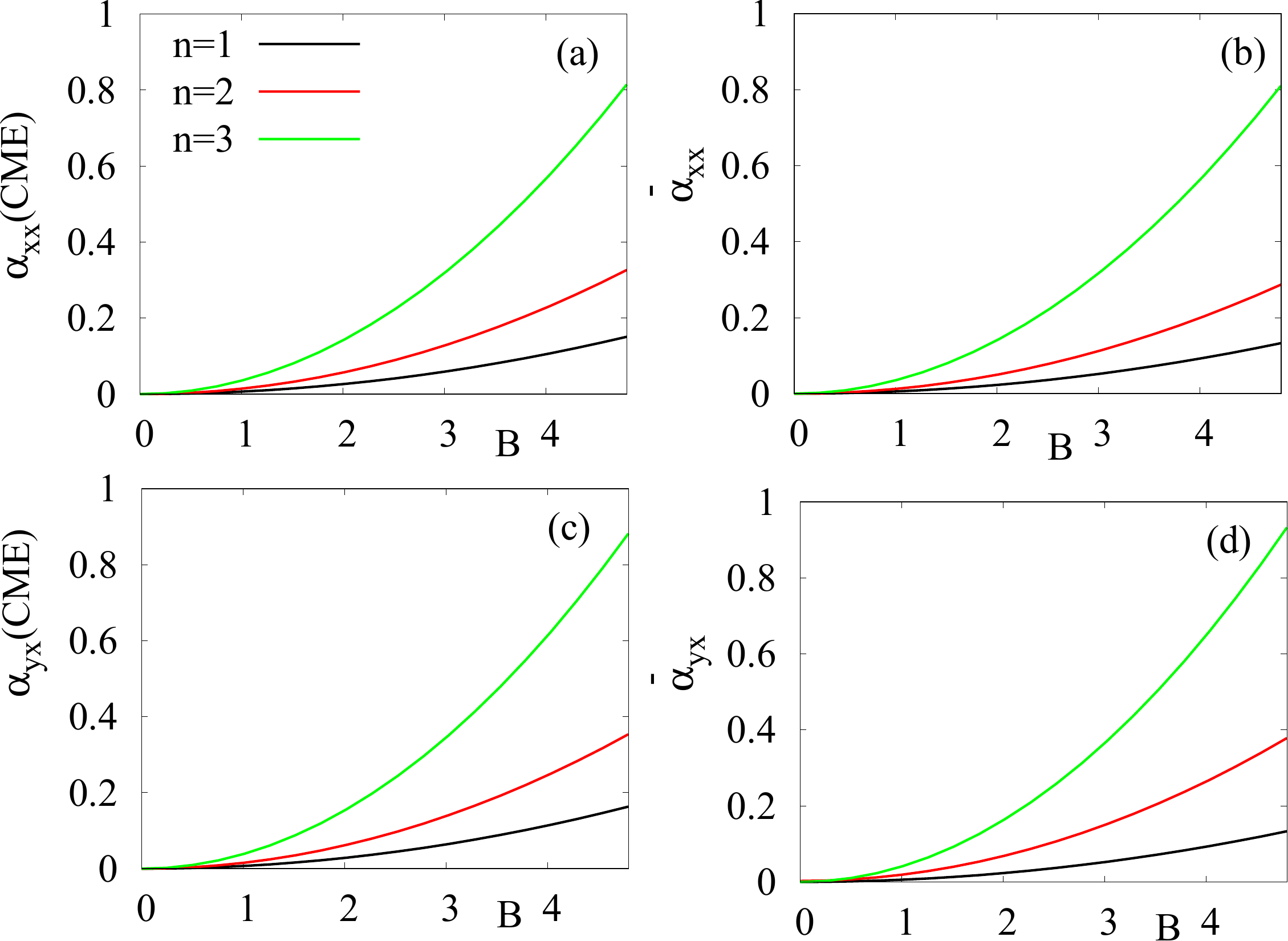} 
\end{center}
\caption{(Color online) 
\textcolor{black}{The variation of the dominating term $\alpha_{xx}({\rm CME})$ and  $\alpha_{yx}({\rm CME})$ as a function of $B$ is shown in (a) and (c) respectively. The total LTEC $\bar \alpha_{xx}$, computed using Eq.~(\ref{eq_pn6}), and the total TTEC $\bar \alpha_{yx}$, computed using Eq.~(\ref{eq_pn7}), as a function of $\gamma$ are plotted in (b) and (d), respectively. The $B^2$-dependence is commonly observed for all of the
above cases. 
 The parameters used here are $B=3$ Tesla, $\mu=0.07$ eV and $T=10$ K. }
} \label{PN_B} \end{figure}

\begin{figure}[ht]
\begin{center}
\includegraphics[width=8.5cm]{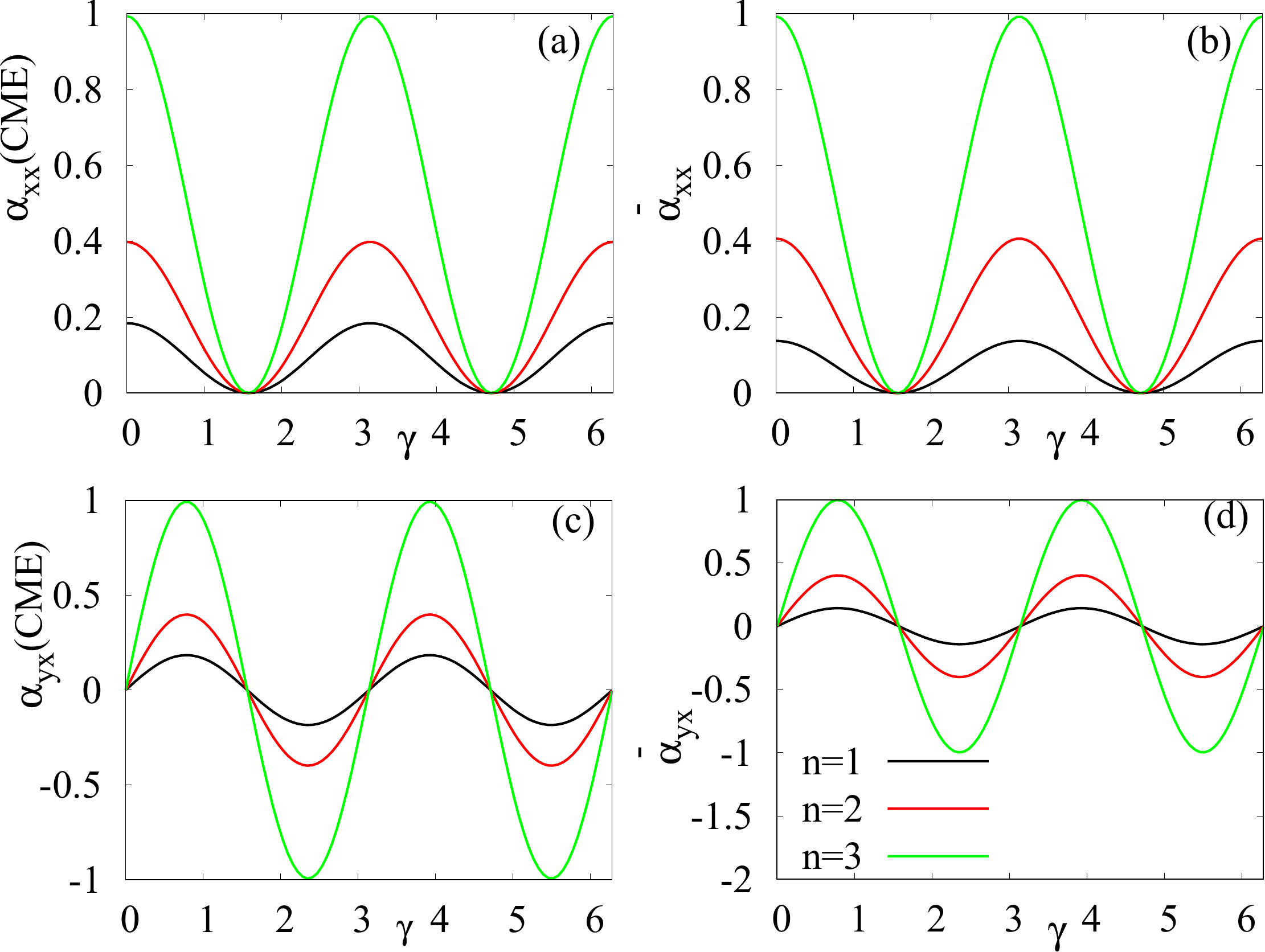} 
\end{center}
\caption{(Color online)\textcolor{black}{The variation of the dominating term $\alpha_{xx}({\rm CME})$ and  $\alpha_{yx}({\rm CME})$ as a function of $\gamma$ at a fixed B value is shown in (a) and (c) respectively. The total LTEC $\bar \alpha_{xx}$, computed using Eq.~(\ref{eq_pn6}), and the total TTEC $\bar \alpha_{yx}$, computed using Eq.~(\ref{eq_pn7}), as a function of $\gamma$ are plotted in (b) and (d), respectively.. Similar to the PHE, 
LTEC goes as $\cos^2 \gamma$ and TTEC varies as $ \sin \gamma \cos \gamma$.  The qualitative behaviors of total TECs remain unaltered as compared to (a) and (c). The parameters used here are $B=3$ Tesla, $\mu=0.07$ eV and $T=10$ K. }
} \label{PN(theta)_ex} \end{figure} 

\subsubsection{Setup 2: Thermo-electric coefficients and  planar Nernst coefficient}
\label{PN_result}

{\bf Magnetic field}: The CME contribution as well as the total contribution of LTEC and TTEC are shown in Fig.~\ref{PN_B}.
We {find that} both LTEC and TTEC vary quadratically with the magnetic field.
This observation  can be verified using the low-energy model (see Eq.~(\ref{eq_xxt_PN}) and Eq.~(\ref{eq_yxt_PN})).
The important point to note here is that {CME is the main origin for the magnetic field dependence of TECs. 
Interestingly, for a given value of 
$B$, ${\bar \al}_{xx}$ and ${\bar \al}_{yx}$ increase non-linearly with $n$ which reflects the multi-Weyl nature in these coefficients.  

{\bf Angle}:
We first plot the CME term of TECs in Fig.~\ref{PN(theta)_ex}, in particular $\al_{xx}$(CME) in 
Fig.~\ref{PN(theta)_ex}(a) and $\al_{yx}$(CME) in Fig.~\ref{PN(theta)_ex}(c), respectively. The numerical findings again 
satisfy the analytical results based on the low-energy model, i.e., 
 $\alpha_{xx}({\rm CME})$ $\propto$ $\cos^2 \gamma$ and $\alpha_{yx}({\rm CME})$ 
 $\propto$ $\sin \gamma \cos \gamma$. The behavior of total TECs are then shown 
 in Fig.~\ref{PN(theta)_ex}(b) and Fig.~\ref{PN(theta)_ex}(d) where both 
${\bar \al}_{xx}$ and ${\bar \al}_{yx}$ exhibit $\cos^2 \gamma$ and
$\sin \gamma \cos \gamma$ {dependence,} respectively. 
The multi-Weyl character is reflected in the non-linear enhancement of the 
amplitude of oscillation for ${\bar \alpha}_{xx}$ and ${\bar \alpha}_{yx}$ with $n$.  

\begin{figure}[ht]
\begin{center}
\includegraphics[width=8.5cm]{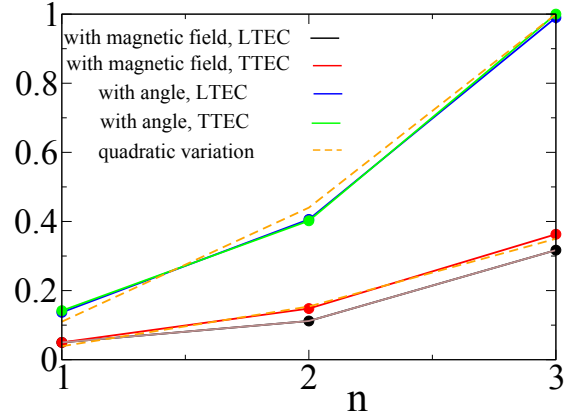} 
\end{center}
\caption{(Color online) We show the topological charge scaling of LTEC and TTEC 
in the planar Nernst setup with the data obtained from 
magnetic field variation and angular variation. The solid black and blue lines
represent the data for LTEC from Fig.~\ref{PN_B}(b) (for $B=3$ Tesla) and 
Fig.~\ref{PN(theta)_ex}(b) (for $\gamma=0$), respectively. The 
solid red and green lines represent the data for TTEC from Fig.~\ref{PN_B}(d) (for $B=3$ Tesla) and 
Fig.~\ref{PN(theta)_ex}(d) (for $\gamma=\pi/2$), respectively. \textcolor{black}{Both the LTEC and the TTEC follow the quadratic variation with $n$ as clearly visible in this figure.}    
} \label{n_PN_ex} \end{figure} 


\begin{figure}[htb]
\centering
\includegraphics[width=8.5cm]{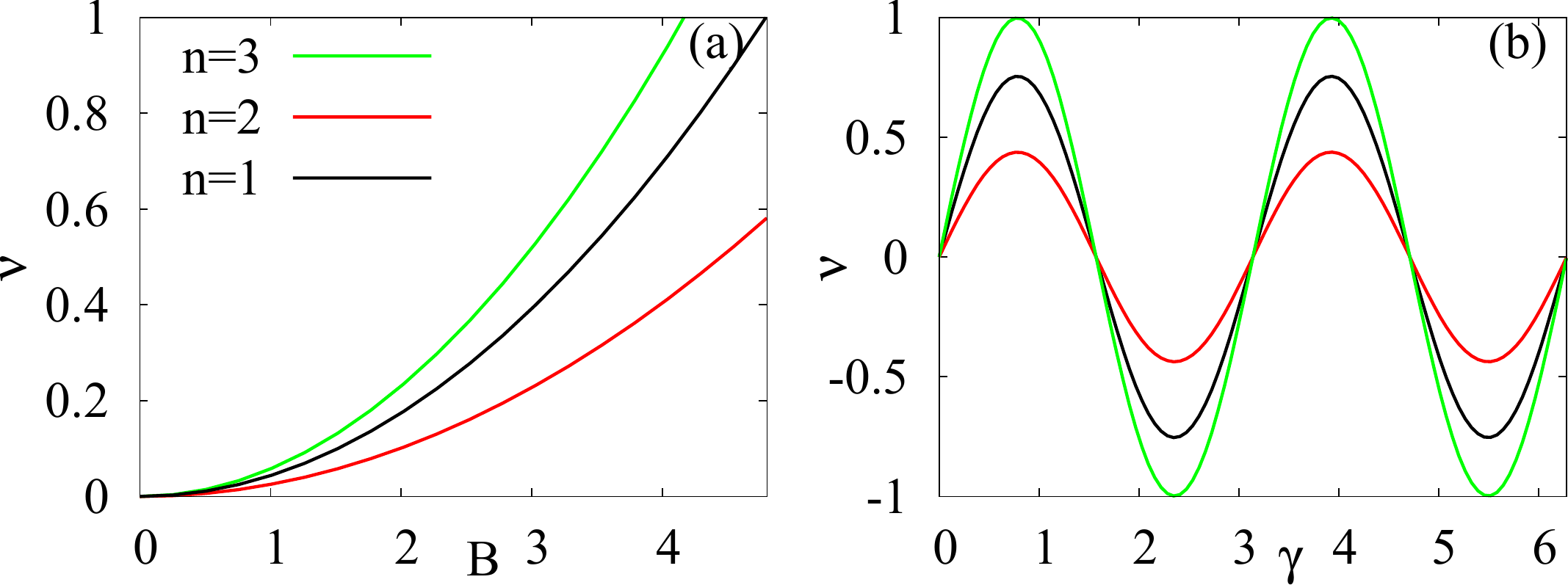} 
\caption{(Color online) The variation of planar Nernst coefficient $\nu$, obtained using Eq.~(\ref{nrnst_eq1}), as a function of (a) magnetic field B and (b) angle $\gamma$ is shown for $n=1,~2$ and $3$. The figure shows that $\nu$ varies as $B^2 \sin \gamma \cos \gamma$. Here, we consider $T=10$ K and $\mu=0.05$ eV. }
\label{nrst_ex_plot}
\end{figure}

{\bf Scaling with $\mathbf{n}$:}
In Fig.~\ref{n_PN_ex}, we show that longitudinal and transverse TECs, $\al_{xx}$ and $\al_{yx}$, 
respectively, vary quadratically with $n$. This reflects the fact that the low-energy model is able to 
capture the underlying physics {of TECs} more quantitatively as compared to LMC and PHC. It is clear that the quadratic dependence {with $n$} of $\al$'s (Fig.~\ref{PH_n}) is more clear than the cubic dependence 
{with $n$} of $\sigma$'s ((Fig.~\ref{n_PN_ex})) for $1\le n\le 2$.
To be precise, the CME term is the main origin for $n^2$ dependence of both TTEC and LTEC while the velocity term, which scales as $n$, contributes sub-dominantly to LTEC in m-WSMs.

{\bf Planar Nernst Coefficient:}
We shall now compute the planar Nernst coefficient $\nu$ using the lattice model of m-WSMs ($n=1$, $2$ and $3$).
We find that $\nu$ varies quadratically with $B$ for all $n$ 
as shown in { Fig.~\ref{nrst_ex_plot}(a).} This behavior is similar to the
behavior of all the transport coefficients in both 
setups. On the other hand, the
 angular dependence of the planar Nernst coefficient appears to be
 $\sin \gamma \cos \gamma $ which is similar with the behavior of
 PHC and TTEC (see Fig.~\ref{nrst_ex_plot}(b)). This behavior is consistent with the analytical result  as given in Eq.~(\ref{nrnst_ana}).
 Interestingly, we find that unlike $\si$ and $\al$, $\nu$ does not exhibit a monotonic behavior with topological charge $n$ at a fixed $B$ and $\gamma$.
 This non-monotonic dependence of PNC on monopole charge can be explained from analytical functional form 
 obtained from $f_1$ and $f_2$ in Eq.~(\ref{nrnst_ana}).
It is also clear from the Fig.~\ref{nrst_ex_plot} that the multi-Weyl nature is clearly reflected in the PNC since $\nu$ for $n=1$ is distinctly different from double and triple WSMs. 
 \textcolor{black}{We note that PNC is an admixture of $\sigma$, mediated by CA, and thermo-electric coefficients $\alpha$, caused by CME. The dependence of PNC on the external parameters such as ${\bm B}$ and ${\gamma}$ can thus be related to the PHC and TTEC. On the other hand, the anisotropic dispersion imprints its signature on PNC via the topological charge and chemical potential; however, their functional forms  in PNC are different from that of observed in PHC and TTEC.}

\section{Conclusions}
\label{cons}

\textcolor{black}{In this work, we study several intriguing transport properties such as LMC, PHC, TECs and PNC for type-I m-WSMs without tilt,
characterized by the topological charge $n$ being more than unity, using semiclassical Boltzmann transport 
theory with the relaxation time approximation. \textcolor{black}{It is clear that anisotropic non-linear dispersion in m-WSMs causes enhanced transport behavior as compared to the isotropic linear single WSMs.
Interestingly, the non-uniform responses, depending on the orientation of the applied fields, can in general be obtained due to the anisotropy.  Furthermore,
there exist more number of conducting Fermi arc states in m-WSMs than the single WSMs leading to enhanced transport in m-WSMs.}
We here mainly focus 
on the co-planar setups where external magnetic field ${\mathbf B}$ and electric field ${\mathbf E}$ or temperature gradient
$ \nabla {\mathbf T}$ lie in the same plane. Using the ${\mathbf E}$-${\mathbf B}$ arrangement, one
can observe {electric coefficients $\sigma$'s such as PHE and LMC} while 
$\nabla {\mathbf T}$-${\mathbf B}$ setup is considered for the  measurement of TECs $\alpha$'s and subsequently 
PNC can be investigated using $\sigma$ and $\alpha$. We validate our low-energy model based analytical results through the numerical lattice calculations. We emphasize
that our work can stimulate experimental efforts to uncover the multi-Weyl nature, specially, the monopole charge dependence of
different transport coefficients as the setup considered here are realizable in experiments\cite{Jia_2016, Xu_2016, Erfu_2016, Li_2018, Liang_2018, Wang_2018, deng19, QLi:2016,Chen_2018, Singha_2018,
kumar18,Watzman_2018,Liang_2017,Rana_2018,Hess_2018}. \textcolor{black}{It is important to note that the transport properties as derived in this work not only become finite but also are expected to show similar magnetic and angular dependencies in an time-reversal symmetric but inversion broken low energy model of m-WSM. However, to predict the correct experimental behavior of the transport properties in an inversion broken m-WSM, one needs to study the lattice model which is an open interesting question and we leave it for future study.} 
At the same time, we note that it would 
be really interesting to investigate these transport properties for 
type-II m-WSM {which we leave for future study}.
}

\textcolor{black}{
In the presence of co-planar electric and magnetic fields, not perfectly aligned with each other
(with $\gamma$ being the angle between ${\mathbf E}$ and ${\mathbf B}$), {we} derive analytical
expressions for both LMC and PHC {using the low-energy model}. 
Interestingly, we find that at zero temperature 
PHC goes as $n^3B^2\sin \gamma \cos \gamma$ whereas LMC follows $n^3B^2 \cos^2 \gamma$ dependence. 
Therefore, it is clear that the magnitude of LMC and PHC both  increase with $n^3$ as we go 
from single WSM $n=1$ to triple WSM $n=3$. 
This is due to the fact that number of conducting channel increases with $n$.
We also find that for single WSM, the magnitude of LMC and PHC  both decrease rapidly as $ \mu^{-2}$ with doping compared
to double and triple WSMs where the magnitude drops as $\mu^{-2/n}$ at $T=0$. On the other hand, at finite
temperature, we show that both LMC and PHC receive a quadratic temperature correction (i.e., scales as $T^2$) with linear
and quadratic topological charge. 
We emphasize that our numerical findings further support that CA is the key ingredient
behind all of these observations. 
 }


\textcolor{black}{Moving on to the thermo-electric responses,
we investigate TECs and PNC for m-WSMs in a setup where co-planar thermal gradient
and magnetic field are not perfectly aligned with each other
(with $\gamma$ being angle between ${\mathbf B}$ and $\mathbf {\nabla T}$).
Interestingly, we find that unlike LMC and PHC, both longitudinal and transverse TECs vary quadratically 
with monopole charge and linearly with temperature.
In particular, the longitudinal TEC varies as $n^2 B^2 T \cos^2 \gamma$ while transverse TEC follows as 
$ n^2 B^2 T \sin \gamma \cos \gamma$. 
Additionally, TECs decay more rapidly with chemical potential {as $\mu^{-2/n-1}$ compared to both} PHC and LMC.
Hence the electric and thermo-electric coefficients have different dependencies on the inherent parameters of m-WSMs.
We clearly show using numerical treatment that  the CME  governs the TECs. Therefore, it is
essential to mention that CME and CA imprint distinct signature on the transport properties in m-WSMs.
Moreover, unlike single WSM, there exists quantitative difference between longitudinal
transport coefficients in 
the case of double and triple WSMs {when an external field ($E/\nabla T$) is applied along the anisotropic 
direction of the underlying energy dispersion.}
Finally, we study PNE which is of a very different nature from the conventional Nernst effect
and even Berry phase mediated anomalous Nernst effect. We find that although PNC behaves qualitatively 
in an identical manner with ${\mathbf B}$ and $\gamma$ 
as compared to the transverse transport coefficient (PHC and TTEC), it does not exhibit a monotonic variation with $n$ like all the other transport coefficients. \textcolor{black}{Therefore, by looking at the scaling of transport coefficients with the monopole charge, which can be experimentally verifiable, one can distinguish a double and triple WSMs from a single WSM.}}

\textcolor{black}{\textit{Note added:} During the completion of our work, we came across the paper~\cite{Sharma_2020}, which
discusses the planar Nernst coefficient for Dirac and single Weyl
semimetals.}

 \begin{acknowledgements}
  We sincerely thank Renato M. A. Dantas for fruitful discussions.
 \end{acknowledgements}

\newpage
\appendix

\begin{widetext}

\section{Calculational detail of Eq.~(\ref{eq_ehc}) in planar Hall setup}
\label{formula_PHC}

\textcolor{black}{
Here our aim is to achieve a modified distribution function.
 We can start from the Eq.~\ref{eq_BZf} and using $\mathbf{\dot{r}}$ (Eq.~(\ref{eq_motion1})) and $\mathbf{\dot{k}}$
(Eq.~(\ref{eq_motion2}) ), one can obtain
\be
e E v_x + e^2 B E \cos \gamma (\mathbf{v_k}.\mathbf{\Omega_k}) 
\frac{\partial f_{0}}{\partial \epsilon} + eB (-v_z \sin \gamma
\frac{\partial}{\partial k_x} + (v_x \sin \gamma -v_y \cos \gamma )
\frac{\partial}{\partial k_z} +v_z \cos \gamma \frac{\partial}{\partial k_y}
) f_{k} = \frac{f_{0} -f_k}{D \tau}
\label{eq1}
\ee
where $D(\mathbf{B},\mathbf{\Omega_k} )= (1+ e(\mathbf{B}.\mathbf{\Omega_k}))^{-1}$ is the phase 
factor. Now the ansatz we are following is given below
\be
f_k-f_{0}= (e D E \tau v_x + e^2 DBE \tau \cos \gamma (\mathbf{v_k}.\mathbf{\Omega_k})
+ \mathbf{v.\Gamma })
\label{eq2}
\ee
where $\Gamma$ is the correction factor due to magnetic field $\mathbf{B}$.
 Therefore, Eq.~(\ref{eq1}) takes the form  
 \be eB
 (-v_z \sin \gamma
\frac{\partial}{\partial k_x} + (v_x \sin \gamma -v_y \cos \gamma )
\frac{\partial}{\partial k_z} +v_z \cos \gamma \frac{\partial}{\partial k_y}
) + (e E D \tau (v_x + eB \cos \gamma (\mathbf{v_k.\Omega_k}) + 
\mathbf{v.\Gamma} )  ) =\frac{\mathbf{v.\Gamma} }{D \tau}
\label{eq3}
 \ee
 Hence,  the distribution function $f_k$ becomes 
 \be
 f_k= f_{0} - e D E \tau (v_x + eB \cos \gamma (\mathbf{v_k.\Omega_k} ))
 \frac{\partial f_{0}}{\partial \epsilon} -
 eDE \tau (v_x c_x \sin \gamma + v_y c_y \cos \gamma + v_z c_z)
  \frac{\partial f_{0}}{\partial \epsilon}
 \ee
 where $c_x$, $c_y$ and $c_z$ are correlation factors which incorporate 
 Berry phase effects and related $\Gamma$. 
 Therefore, the general expression of the PHC in the above configuration from the semiclassical Boltzmann equation can be written
as~\cite{Nandy_2017, Nandy_2018, Sharma:2016}
\ba
\sigma_{yx}&=& e^{2}\int\frac{d^{3}k}{(2\pi)^{3}}D\tau\left(-\frac{\partial f_{0}}{\partial \epsilon}\right) 
[(v_{y}+\frac{eB\sin \gamma}{\hbar}(\mathbf{v_{k}}\cdot\mathbf{\Omega_{k}})) 
(v_{x}+\frac{eB\cos \gamma}{\hbar}(\mathbf{v_{k}}\cdot\mathbf{\Omega_{k}}))]\nonumber \\
&=&{L^{11}_{yx}}
-e^2\int \frac{d^{3}k}{(2\pi)^{3}} \mathbf{\Omega}_z f_{0}
e^2 \tau \int \frac{d^{3}k}{(2\pi)^{3}} (c_x v_x \sin \gamma + 
c_y v_y \cos \gamma +c_z v_z )v_y (-\frac{\partial f_{0}}{\partial \epsilon})
\label{eq4} 
\ea 
with $c_x=\Gamma_x/eE \tau \sin \gamma$, $c_y= \Gamma_y/eE \tau \cos \gamma$ and 
$c_z=\Gamma_z/eE \tau $. Before that we define $m_{ij}= \partial^2 H/\partial k_i \partial k_j $. 
The detail expression for 
$\Gamma_{x,y,z}$ are given below:
\ba
\Gamma_x&=&\frac{\sin \gamma (NM_3 + \Gamma_z \frac{eB}{m_{zz}} )}{M_2}\non\\
\Gamma_y & =& -\frac{\cos \gamma (NM_3+ \Gamma_z \frac{eB}{m_{zz}} ) }{M_2}\non \\
\Gamma_z &=& \frac{N (M_1 M_2 + M_3 M_4)}{ \frac{1}{D^2\tau^2} -(\frac{eB \cos\gamma}{m_{yz}} 
-\frac{eB \sin\gamma}{m_{xz}}  )^2 -\frac{eB M_4}{m_{zz}} }
\ea
and 
 $N=e^2 EB D \tau$, $M_1=-\sin \gamma/ m_{xx} + \cos \gamma/ m_{xy} + eB \cos \gamma (R \cos \gamma -P \sin \gamma)  $
 , $M_2= 1/D\tau -eB \sin \gamma/ m_{xz} + eB \cos \gamma/m_{yz} $, $M_3=eB \cos \gamma T + 1/m_{xz}$ and 
 $M_4=eB \sin 2\gamma/m_{xy} -eB \cos^2 \gamma/m_{yy} -eB \sin^2 \gamma/m_{xx} $ with 
 $P=\sigma_x/m_{xx}+ \sigma_y/m_{xy} + \sigma_z/m_{xz} $, $R=\sigma_x/m_{xy}+ \sigma_y/m_{yy} + \sigma_z/m_{yz}$
 and $P=\sigma_x/m_{xz}+ \sigma_y/m_{yz} + \sigma_z/m_{zz}$.
Numerical calculation shows that $c_x,c_y,c_z \to 0$. Therefore, the modified 
 distribution function would be simply given by 
 \be
 f_k= f_{0} - e D E \tau (v_x + eB \cos \gamma (\mathbf{v_k.\Omega_k} ))
 \frac{\partial f_{0}}{\partial \epsilon}
 \ee
 which we use in our analysis in computing $\si$ and $\al$.}

\section{Calculation of LMC in regular setup}
\label{app1}

Now we shall compute the LMC and electrical Hall conductivity for the continuum model (\ref{eq_multi1}). 
We present this calculation to clearly mention the calculation details which we follow for Sec.~\ref{app2}, \ref{app3}.
Here we assume the electric and magnetic field to have the following form: $\mathbf{E}=E\hat{j} $ and 
$\mathbf{B}=B\hat{j}$. we refer $\frac{\partial f_{0}}{\partial \epsilon}=\tilde{f}_{0}$.
This is the coefficient of electric charge current along $j$ direction for an applied electric field 
in $j$ direction:  $\sigma_{jj}$ 
\be
\sigma_{jj}=e^{2}\int\frac{d^{3}k}{(2\pi)^{3}}\tau D[({v_{j}}+
\frac{eB_j}{\hbar}(\mathbf{v_{k}}\cdot\mathbf{\Omega_{k}}))^{2} 
]\left(-\frac{\partial f_{0}}{\partial \epsilon}\right) 
\label{eq_cm1}
\ee
We now decompose the above expression term by term to investigate it more rigorously: $\sigma_{jj}=
\sigma_{jj}^{(1)}+\sigma_{jj}^{(2)}+2\sigma_{jj}^{(3)}$ where 
\ba
\sigma ^{(1)}_{jj} &=& \tau e^2  \int \frac{d^3\mathbf{k}}{(2 \pi)^3}  
\frac{ (\mathbf{v}_j)^2 }{1+e B \Omega_j/\hbar  } 
\left(- \frac{\partial  f_0}{\partial \epsilon} \right) \label{eq_cm2},\\
\sigma ^{(2)}_{jj} &=& \tau e^4 \frac{B_j^2}{\hbar^2} \int \frac{d^3\mathbf{k}}{(2 \pi)^3} 
\frac{ ( \mathbf{\Omega}_{\mathbf{k}} \cdot \mathbf{v}_{\mathbf{k}})^2}{1+e B \Omega_j/\hbar}
\left(- \frac{\partial  f_0}{\partial \epsilon} \right) \label{eq_cm3}, \\
 \sigma ^{(3)}_{jj} &=& \tau e^3 \frac{B_j}{\hbar} \int \frac{d^3\mathbf{k}}{(2 \pi)^3} 
 \frac{ (\mathbf{v}_j) ( \mathbf{\Omega}_{\mathbf{k}} \cdot \mathbf{v}_{\mathbf{k}})}
 {1+e B \Omega_j/\hbar} 
 \left(- \frac{\partial  f_0}{\partial \epsilon} \right)  \label{eq_cm4},\\
 \ea
We note here that two LMCs are given by $\si_{xx}$ and $\si_{zz}$, $J_x=\si_{xx} E_x$ and
$J_z=\si_{zz} E_z$.

We make resort to cylindrical polar co-ordinate to do the analytical calculation. $\int {d^3\mathbf{k}}/{(2 \pi)^3}=
(1/(2 \pi)^3)\int^{\infty}_{0} k_{\bot} dk_{\bot} \int_{-\infty}^{\infty} dk_z \int^{2\pi}_{0} d \phi$. 
We need to compute the following momentum integral for finite temperature; we use the Sommerfeld expansion.
\be
 \tilde {f}(\epsilon)_{0}= \beta \int^{\infty}_0 (\frac{h(\epsilon)}{1+e^{\beta(\epsilon-\mu)}}-
 \frac{h(\epsilon)}
 {(1+e^{\beta(\epsilon-\mu)})^2}) d\epsilon
\ee
We use the change of variable $\beta(\epsilon-\mu)=x$ and above integral becomes
\ba
 \tilde {f}(\epsilon)_{0}&=&\int^{0}_{-\beta \mu} (\frac{h(\mu+x/\beta)}{1+e^{x}}-\frac{h(\mu+x/\beta)}
 {(1+e^{x})^2}) dx \non \\
 &+& \int^{\infty}_{0} (\frac{h(\mu+x/\beta)}{1+e^{x}}-\frac{h(\mu+x/\beta)}
 {(1+e^{x})^2}) dx 
\ea
In the first integral we use $x\to -x$ and using the fact  that $(e^{-x}+1)^{-1}=1-(e^x+1)^{-1}$.
We assume $\beta\mu\gg 1$ and obtain 
\be
\tilde {f}(\epsilon)_{0}=\int^{\infty}_{0} (\frac{1}{1+e^{x}}-\frac{1}
 {(1+e^{x})^2})(h(\mu-x/\beta)+h(\mu+x/\beta)) dx
\ee
Now one can expand $h$ around $\mu$ as the integrand decreases exponentially with increasing
$x$
\be
\tilde {f}(\epsilon)_{0}=\int^{\infty}_{0} (\frac{1}{1+e^{x}}-\frac{1}
 {(1+e^{x})^2})(2 h(\mu)+ K_B^2 T^2 x^2 h''(\mu)) 
\ee
In our case, $h(k)=(k-\mu)k^\nu$ and $\tilde {f}(\epsilon)_{eq}=\pi^2 K_B^2 T^2\nu \mu^{\nu-1}$ and 
when $h(k)=k^\nu$ then $\tilde {f}(\epsilon)_{0}=\mu^{\nu}+\pi^2 K_B^2 T^2\nu(\nu-1) \mu^{\nu-2}$. 


We shall derive the analytical form of LMC in finite temperature by considering 
$-\frac{\partial f_{0}}{\partial \epsilon}=\beta f_{0}(1-f_{0})=
\tilde {f}(\epsilon)_{0}$. 
 \be
\sigma ^{(1)}_{zz} =
\frac{\tau e^2}{(2 \pi)^3} \int^{\infty}_{0}  d k_{\bot} \int_{-\infty}^{\infty} dk_z \int^{2\pi}_{0} d \phi
\frac{k_{\bot} v^4 k_z^2/\epsilon^2}{1+e n^2 v \al_n^2 k_{\bot}^{2n-2}B k_z/(2\epsilon^3 \hbar )} 
\tilde {f}(\epsilon)_{eq} \nonumber
\ee
Now we perform the variable substitution $k_z \rightarrow k_z/v $ and $k_{\bot} \rightarrow k_{\bot}   \alpha_{n}^{-1/n} $
Hence the energy becomes $\epsilon \to \epsilon'=\sqrt{ k_{\bot}^{2n}+k^2_z}$.
\be
\sigma^{(1)}_{zz} =
\frac{\tau e^2}{(2 \pi)^2} \int^{\infty}_{0}  d k_{\bot} \int_{-\infty}^{\infty} dk_z
\frac{k_{\bot} \al_n^{-2/n} v k_z^2/\epsilon'^2}{1+e n^2\al_n^{2/n} k_{\bot}^{2n-2}B k_z/(2\epsilon'^3 \hbar )} 
\tilde {f}(\epsilon')_{0} \nonumber
\ee
We then use another change of variable $k_{\bot} \to {k}_{\bot} ^{1/n}$ and $\epsilon' \to \epsilon''=\sqrt{k_{\bot}^2+k_z^2}$.
\be
\sigma^{(1)}_{zz} =
\frac{\tau e^2}{(2 \pi)^2} \int^{\infty}_{0}  d k_{\bot} \int_{-\infty}^{\infty} dk_z
\frac{k_{\bot}^{2/n-1} \al_n^{-2/n} v k_z^2/(n \epsilon''^2)}{1+e n^2\al_n^{2/n} k_{\bot}^{2-2/n}B k_z/(2\epsilon''^3 \hbar )} 
\tilde {f}(\epsilon'')_{0} \nonumber
\ee
Finally, one can perform another transformation $k_{\bot}=k \sin{\theta}$ and $k_z=k \cos{\theta}$ and 
hence $\epsilon''\to \epsilon'''=k$. $\int^{\infty}_{0}  d k_{\bot} \int_{-\infty}^{\infty} dk_z \to 
\int^{\pi}_0 d\theta \int^{\infty}_0 dk$. 
\ba
\sigma^{(1)}_{zz} &=&
\frac{v \tau e^2\alpha^{-2/n}_n}{n(2 \pi)^2} \int^{\infty}_{0}  dk \int^{\pi}_{0} d\theta
\frac{k^{2/n} \cos^2 \theta (\sin \theta)^{2/n-1}}{1+eB n^2\al_n^{2/n} k^{-2/n} (\sin \theta)^{2-2/n}\cos \theta/(2 \hbar )} 
\tilde {f}(k)_{0} \nonumber \\
&=&
\frac{ v \tau e^2\alpha^{-2/n}_n}{n(2 \pi)^2}  \int^{\pi}_{0} d\theta \int^{\infty}_0 dk
\frac{ \cos^2 \theta (\sin \theta)^{2/n-1}}{1+eB n^2\al_n^{2/n} k^{-2/n} (\sin \theta)^{2-2/n}\cos \theta/(2 \hbar )} 
k^{2/n}\tilde {f}(k)_{0}
\nonumber
\ea
In order to evaluate the integrals we need to perform a series expansion in terms of $eB/\hbar \mu^{2/n}$. We 
use the series expansion $(1+x)^{-1}=\sum_{i=0}^{\infty}(-x)^i$ with $x\ll 1$ for the denominator. Therefore, the 
integral becomes
\be
\sigma^{(1)}_{zz} =
\frac{ v \tau e^2\alpha^{-2/n}_n}{n(2 \pi)^2} \sum_i (-e B n^2 \al_n^{2/n} /2\hbar)^i
\int^{\pi}_{0} d\theta 
 (\cos \theta)^{2+i} (\sin \theta)^{2/n-1+i(2-2/n)} \int^{\infty}_0 dk k^{2/n}k^{-2i/n}\tilde {f}(k)_{0}
 \nonumber
\ee
The leading order terms are given by
\ba
\sigma^{(1)}_{zz} &\simeq&
\frac{ v \tau e^2\alpha^{-2/n}_n}{n(2 \pi)^2} 
 (\frac{\Gamma(1/n)\Gamma(3/2)}{\Gamma(3/2+1/n)} (\mu^{2/n}+ 
 \frac{ \pi^2 K_B^2 T^2 2(2-n) \mu^{2/n-2}}{6 n^2 }) \non \\
&+&  \frac{ e^2 B^2 n^4 \alpha_n^{4/n}}{\hbar^2} \frac{\Gamma(2-1/n)\Gamma(5/2)}{4 \Gamma(9/2-1/n)}
(\mu^{-2/n}+ \frac{\pi^2 K_B^2 T^2 2(2+n) \mu^{-2/n-2}}{6 n^2} ) )
 \label{eq_zz1t}
\ea

Similarly, $\sigma^{(2)}_{zz}$ is given by
\ba
\sigma^{(2)}_{zz} &=&
\frac{ \tau e^4 B^2 n^3 \alpha^{2/n}_n}{4\hbar^2(2 \pi)^2} \sum_i (-e B n^2 \al_n^{2/n} /2\hbar)^i
\int^{\pi}_{0} d\theta 
 (\cos \theta)^{i} (\sin \theta)^{3-2/n+i(2-2/n)} \int^{\infty}_0 dk k^{-2/n}k^{-2i/n}\tilde {f}(k)_{0}
 \nonumber \\
 &\simeq&  \frac{ \tau e^4 B^2 n^3 \alpha^{2/n}_n}{4\hbar^2(2 \pi)^2} 
 \frac{\Gamma(2-1/n)\Gamma(1/2)}{\Gamma(5/2-1/n)} (\mu^{-2/n}+ \frac{\pi^2 K_B^2 T^22(2+n)\mu^{-2/n-2}}{6 n^2})
\label{eq_zz2t}
 \ea
Similarly, $\sigma^{(3)}_{zz}$ is given by
\ba
\sigma^{(3)}_{zz} &=&
\frac{ \tau e^3 B n v }{2 \hbar^2(2 \pi)^2} \sum_i (-e B n^2 \al_n^{2/n} /2\hbar)^i
\int^{\pi}_{0} d\theta 
 (\cos \theta)^{i+1} (\sin \theta)^{1+i(2-2/n)} \int^{\infty}_0 dk k^{-2i/n}\tilde {f}(k)_{0}
 \nonumber \\
 &\simeq&  -\frac{ \tau e^4 B^2 n^3 v\alpha^{2/n}_n}{2\hbar^2(2 \pi)^2} 
 \frac{\Gamma(2-1/n)\Gamma(3/2)}{\Gamma(7/2-1/n)}(\mu^{-2/n}+\frac{\pi^2 K_B^2 T^2 2(2+n)\mu^{-2/n-2}}{6n^2})
\label{eq_zz3t}
 \ea
Therefore, the CA term in $\si_{zz}$ is proportional to $B^2  
  (n^3\mu^{-2/n}+n^2 T^2 \mu^{-2-2/n})$. The complete $\si_{zz}$ is also proportional to $ \mu^{2/n}/n + T^2/n^3+ B^2  (n^3\mu^{-2/n}+n^2 T^2 \mu^{-2-2/n} )$.

 Similarly, $\sigma^{(1)}_{xx}$ is given by
\ba
\sigma ^{(1)}_{xx} &=&
\frac{ \tau e^2 n}{v(2 \pi)^3} \sum_i (-e B n v\al_n^{1/n} /2\hbar)^i
\int^{\pi}_{0} d\theta \int^{2\pi}_{0}
 (\cos \phi)^{2+i} (\sin \theta)^{3+i(2-1/n)} \int^{\infty}_0 dk k^{2-i(1+1/n)}\tilde {f}(k)_{0}
 \nonumber \\
 &\simeq&  \frac{\tau e^2 n}{v(2 \pi)^3} ( \frac{8 \pi }{5} (\mu^{2} + \frac{\pi^2 K_B^2 T^2}{3})\non \\
 &+& 
 (e B n v\al_n^{1/n} \mu^{-1-1/n}/2)^2 \frac{3 \pi \Gamma(4-1/n)\Gamma(1/2)}{4 \Gamma(9/2-1/n)} (\mu^{-2/n} + \frac{\pi^2 K_B^2 T^2 2(2+n)\mu^{-2-2/n}}{6n^2}) )
 \label{eq_xx1t}
 \ea 
Similarly, $\sigma^{(2)}_{xx}$ is given by  
\ba
\sigma ^{(2)}_{xx} &=&
\frac{ v \tau e^4 n^3 \al^{2/n}_n B^2}{4\hbar^2(2 \pi)^3} \sum_i (-e B n v\al_n^{1/n}/2\hbar)^i
\int^{\pi}_{0} d\theta \int^{2\pi}_{0} d\phi
 (\cos \phi)^{i} (\sin \theta)^{3-2/n+i(2-1/n)} \int^{\infty}_0 dk k^{-2/n-i(1+1/n)}\tilde {f}(k)_{0}
 \nonumber \\
 &\simeq& \frac{ v \tau e^4 n^3 \al^{2/n}_n B^2 \pi^{3/2}}{2\hbar^2(2 \pi)^3}\frac{\Gamma(2-1/n)}{\Gamma(5/2-1/n)} 
 (\mu^{-2/n}+ \frac{\pi^2 K_B^2 T^2 2(2+n)\mu^{-2-2/n}}{6 n^2})
 \label{eq_xx2t}
 \ea
 
Similarly, $\sigma^{(3)}_{xx}$ is given by  
\ba
\sigma ^{(3)}_{xx} &=&
\frac{ \tau e^3 n^2 \al^{1/n}_n B}{2\hbar^2(2 \pi)^3} \sum_i (-e B n v\al_n^{1/n} /2\hbar)^i
\int^{\pi}_{0} d\theta \int^{2\pi}_{0}
 (\cos \phi)^{1+i} (\sin \theta)^{3-1/n+i(2-1/n)} \int^{\infty}_0 dk k^{1-1/n-i(1+1/n)}\tilde {f}(k)_{0}
 \nonumber \\
 &\simeq& -\frac{ v \tau e^4 n^3 \al^{2/n}_n B^2 \pi}{2\hbar^2(2 \pi)^3}\frac{\Gamma(3-1/n)\Gamma(3/2)}{\Gamma(7/2-1/n)} 
(\mu^{-2/n}+ \frac{\pi^2 K_B^2 T^2 2(2+n)\mu^{-2-2/n}}{6 n^2})
 \label{eq_xx3}
 \ea 
 
 Therefore, the CA term in $\si_{xx}$ is proportional to $B^2  
  (n^3\mu^{-2/n}+n^2 T^2 \mu^{-2-2/n})$. The complete $\si_{xx}$ is also proportional to $n \mu^2 + n T^2 +B^2  (n^3\mu^{-2/n}+n^2 T^2 \mu^{-2-2/n} )$.
 
 \section{Calculations for LMC, PHC, TECs and PNC in coplanar setup}
\label{app2}

 Having discussed the LMC in normal regular setup, we shall now turn our attention to PH setup. Here we compute two main quantities 
$\si_{jj}$ and $\si_{ij}$. The magnetic field here is assumed to have the form: $\mathbf{B}=
B\cos\gamma\hat{j}+B\sin\gamma\hat{y}$ and $\mathbf{E}=E\hat{j}$ with $j=x,z$. Now using the same procedure mentioned in Appendix~\ref{app1}, the CA term in $\si_{xx}$ is proportional to $B^2  \cos^2 \gamma
  (n^3\mu^{-2/n}+n^2 T^2 \mu^{-2-2/n})$. The complete $\si_{xx}$ is proportional to $ n \mu^2 + n T^2
+  B^2  \cos^2 \gamma  (n^3\mu^{-2/n}+n^2 T^2 \mu^{-2-2/n} )$.  
On the other hand, the CA term in $\si_{zz}$ is proportional to $B^2  \cos^2 \gamma
  (n^3\mu^{-2/n}+n T^2 \mu^{-2-2/n})$. The complete $\si_{zz}$ is also proportional to $
\mu^{2/n}/n + T^2 \mu^{2/n-2}/n^3+   B^2  \cos^2 \gamma  (n^3\mu^{-2/n}+n^2 T^2 \mu^{-2-2/n} )$.
Similarly, we find that the CA term in $\sigma_{yx}$ is proportional to $B^2  \cos \gamma\sin \gamma
  (n^3\mu^{-2/n}+n T^2 \mu^{-2-2/n})$. The complete $\sigma_{yx}$ is also 
  proportional to $B^2  \cos \gamma\sin \gamma  (n^3\mu^{-2/n}+n T^2 \mu^{-2-2/n} )$.
 Now, the CA term in $\si_{yz}$ is proportional to $ B^2 
\cos \gamma\sin \gamma (n^3 \mu^{-2/n}+ n T^2 \mu^{-2-2/n})$. The complete $\si_{yz}$ is proportional 
to $ B^2  \cos \gamma\sin \gamma (n^3 \mu^{2-2/n}+ n T^2 \mu^{-2/n}+n^3\mu^{-2/n}+n T^2 \mu^{-2-2/n})$.
 We note that $\gamma=0$  corresponds to the regular setup where 
${\mathbf B}$ and ${\mathbf E}$ lie in two different planes. The above results reduce to the regular result
for LMC and PHC if $\gamma$
is set to zero.

We shall now present the thermo-electrical coefficients for transport. The magnetic field here is assumed to have the same form as mentioned above: $\mathbf{B}=
B\cos\gamma\hat{j}+B\sin\gamma\hat{y}$ and $\mathbf{\nabla T}=\nabla T\hat{j}$ with $j=x,z$. 
Therefore, the CME term in $\al_{xx}$ is proportional to
$n^2 B^2 T \cos^2 \gamma \mu^{-1-2/n}$. The complete $\al_{xx}$ is proportional 
to $T(-\mu n +B^2 \mu^{-1-2/n}n^2 \cos^2 \gamma)$.
  
On the other hand, one can find that the CME term in $\al_{zz}$ is proportional to
  $n^2 B^2 T \cos^2 \gamma \mu^{-1-2/n}$. The complete $\al_{zz}$ is proportional to 
  $T(-\mu^{2/n-1} /n^2 +B^2 \mu^{-1-2/n}n^2 \cos^2 \gamma)$. 
We also find that the CME term and complete $\al_{yx}$   are both proportional
  to $T B^2 n^2 \cos \gamma \sin \gamma\mu^{-1-2/n}$. 
Interestingly, the CME tern in $\al_{yz}$ is proportional to $T B^2 n^2 \cos \gamma \sin \gamma \mu^{-1-2/n}$ whereas the complete 
 $\al_{yz} \propto n^2 TB^2 \cos \gamma \sin \gamma (\mu^{-1-2/n}+ \mu^{1-2/n})$.
 We would like to point out that $\gamma=0$  corresponds to the regular setup where 
${\mathbf B}$ and ${\mathbf \nabla T}$ lie in two different planes. The above results reduce to the regular result
for LTEC and TTEC if $\gamma$
is set to zero. After the full calculation, one can obtain the Nernst coefficient to be 
\be
 \tiny
  \nu \sim \frac{ O(B^2 n^2 T \mu^{-1-2/n}) \cos \gamma \sin \gamma (O(B^0 n T^2) + O(B^2 n^3 \mu^{-2/n}) \cos^2 \gamma) -
  O(B^2 n^3  \mu^{-2/n}) \cos \gamma \sin \gamma (O(B^0 T n)- O(B^2 n^2 T \mu^{-1-2/n} \cos^2 \gamma  ))}
  {(O(B^0 n T^2) + O(B^2 n^3) \cos^2 \gamma)^2 + O(B^4 n^6  \mu^{-4/n}) \cos^2 \gamma \sin^2 \gamma }
  \label{nrnst_ana_appdx}
  \ee

 \end{widetext}

\end{document}